\documentclass[twocolumn,prd,aps,superscriptaddress,preprintnumbers,tightenlines,showpacs,nofootinbib,eqsecnum,amsfonts,amsmath]{revtex4-1}
\usepackage[normalem]{ulem}
\usepackage{placeins}
\usepackage{color}
\usepackage{calc}
\usepackage{amsmath,amssymb,graphicx}
\usepackage{tensor}
\usepackage{bm}
\usepackage{times}
\usepackage[varg]{txfonts}
\usepackage[colorlinks, pdfborder={0 0 0}]{hyperref}
\usepackage{float}
\usepackage{dcolumn}
\usepackage[nolist,nohyperlinks]{acronym}
\usepackage{xspace}
\usepackage[abs]{overpic}
\usepackage{pict2e}
\usepackage{enumitem}
\usepackage[usenames,dvipsnames]{xcolor}
\usepackage[utf8]{inputenc}
\usepackage{acronym}
\usepackage{gensymb}
\usepackage{cleveref}
\usepackage[normalem]{ulem}
\usepackage{longtable}
\usepackage{verbatim}
\usepackage{multirow}

\usepackage{subfigure}
\usepackage{placeins}

\newcommand{\bsub}{\begin{subequations}}
\newcommand{\esub}{\end{subequations}}

\definecolor{frenchlila}{rgb}{0.53, 0.38, 0.56}

\newcommand{\rhod}{\dot{\rho} }

\begin{document}

 \title{Modeling gravitational waves from exotic compact objects}

\author{Alexandre Toubiana}
\affiliation{APC, AstroParticule et Cosmologie, 
Université de Paris, CNRS, Astroparticule et Cosmologie, F-75013 Paris, France}
\affiliation{Institut d'Astrophysique de Paris, CNRS \& Sorbonne
 Universit\'es, UMR 7095, 98 bis bd Arago, 75014 Paris, France}

 \author{Stanislav Babak}
\affiliation{APC, AstroParticule et Cosmologie, 
Université de Paris, CNRS, Astroparticule et Cosmologie, F-75013 Paris, France}
\affiliation{Moscow Institute of Physics and Technology, Dolgoprudny, Moscow region, Russia}

\author{Enrico Barausse}
\affiliation{SISSA, Via Bonomea 265, 34136 Trieste, Italy and INFN Sezione di Trieste}
\affiliation{IFPU - Institute for Fundamental Physics of the Universe, Via Beirut 2, 34014 Trieste, Italy}

\author{Luis Lehner}
\affiliation{Perimeter Institute, 31 Caroline St N, Ontario, Canada}

\begin{abstract}
Exotic compact objects can be difficult 
to distinguish from black holes in the inspiral phase of the binaries observed by 
gravitational-wave detectors, but
significant differences may be present in the
merger and post-merger signal.
We introduce a toy model capturing the salient features
of binaries of exotic compact objects with compactness below $0.2$, which do not collapse promptly following the merger. We use it
to assess their detectability with current and future detectors, and whether
they can be distinguished from black hole binaries.
We find that the Einstein Telescope (LISA) could observe exotic binaries with total mass $\mathcal{O}(10^2) \ M_{\odot}$ ($10^4-10^6 \ M_{\odot}$), and potentially distinguish them from black hole binaries, throughout the observable Universe, as compared to $z\lesssim 1$ for Advanced LIGO.
Moreover, we show that using standard black hole templates for detection could lead to a loss of up to $60\%$ in the signal-to-noise ratio, greatly reducing our chances of observing these signals. Finally, we estimate that if the loudest events in the O1/O2 catalog released by the LIGO/Virgo collaboration were ECO binaries as the ones considered in this paper, they would have left a post-merger signal detectable with model-agnostic searches, making this hypothesis unlikely.
\end{abstract}

\maketitle

\section{Introduction}

In the standard astrophysical paradigm, the only compact objects (with compactness $C=GM/Rc^2\gtrsim 0.1$ where $M$ is the mass of the object and $R$ its radius) are black holes (BHs) and neutron stars (NSs). Among these, BHs are the only ones that can have mass above $3 \ M_{\odot}$. Within this paradigm, extensive efforts have provided theoretical guidance for the search, detection
and analysis of gravitational waves (GWs) from binary black holes (BBHs), binary neutron stars (BNSs) and neutron star–black hole binaries.
To date, detections by the LIGO/Virgo collaboration (LVC) \cite{TheLIGOScientific:2014jea,TheVirgo:2014hva} are in solid agreement
with these predictions~\cite{TheLIGOScientific:2016src,Abbott:2018lct,LIGOScientific:2019fpa,Abbott:2020uma,LIGOScientific:2020stg,Abbott:2020khf,Abbott:2020tfl,Abbott:2020jks}. Nevertheless, the availability of data and its gradually improving accuracy offer compelling motivations for 
exploring possible extensions of this paradigm.
\begin{figure}[th!]
{\centering \includegraphics[scale=0.17]{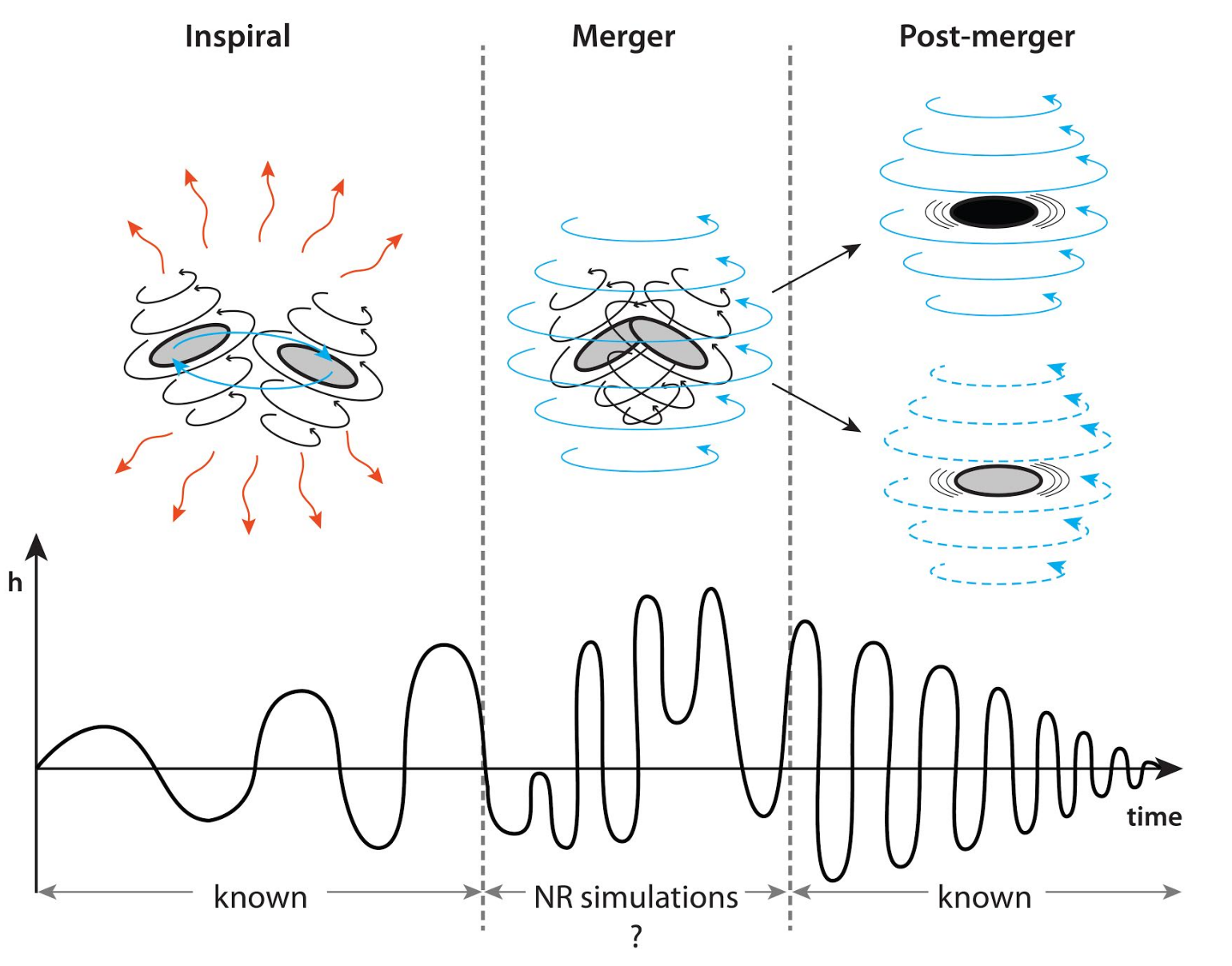}
}
\caption{Schematic representation of our knowledge of the GW signal emitted by an ECO binary: within GR we have a good idea of the qualitative features of the inspiral and post-merger, but the connection between these two regimes is still
largely unknown. We expect that the merger leads to the formation of either a BH or an (excited) object of the same nature as the initial ECOs, possibly rotating. The remnant relaxes through emission of GWs, which are related to its fundamental properties and/or the dynamics of the merger.}\label{scheme_eco}   
 \end{figure}
For instance, extensions of General Relativity (GR) and/or of the Standard Model can give rise to ``exotic compact objects'' (ECOs). Examples of ECOs are boson stars (BSs), see \cite{Liebling:2012fv} and references therein, gravastars \cite{Mazur:2001fv,Visser:2003ge}, wormholes \cite{Visser:1995cc}, fuzzballs \cite{Skenderis:2008qn} and firewalls \cite{Almheiri:2012rt}; see also \cite{saravani2012black,Holdom_2017,Giddings_2014,Abedi_2017} for further possibilities and \cite{Cardoso:2019rvt} for a review. ECOs can have masses ranging from $\mathcal{O}(1 M_{\odot}$) to $\mathcal{O}(10^9 M_{\odot})$ and compactnesses ranging from $0.1 $ to $0.5$ \footnote{We recall that the minimum compactness of stationary BHs in GR is 0.5.}, thus mimicking the gravitational behavior of BHs and NSs to varying degrees. 
For this reason, their identification is potentially 
difficult, requiring the use of specific gravitational wave templates to distinguish them. 
Unfortunately, the vast majority of models of ECOs are not yet complete enough to construct their corresponding 
gravitational wave templates, especially
in the binary merger case, with the exception of (some types of) binary boson stars (BBSs). Furthermore, even for BBSs, full
solutions are available only in a small number of cases \cite{Palenzuela:2017kcg,Bezares:2017mzk,Bezares:2018qwa}. The lack of templates beyond the standard paradigm 
could spoil our chances of observing these systems, because the signal to noise ratio (SNR) is highest with matched filtering techniques. In a sense, the current state of knowledge outside the standard BH/NS paradigm, illustrated in Fig.~\ref{scheme_eco},
resembles the situation in the early 2000s for BBHs within GR\footnote{Fig.~\ref{scheme_eco} is largely inspired from Fig.~9 in \cite{Thorne:2002ys}.}. At that epoch, only the early inspiral and post-merger phases had been reasonably modeled, while no numerical-relativity (NR) simulations for the merger phase were available until~\cite{Pretorius:2005gq,Campanelli:2005dd,Baker:2005vv} and the large body of work since then.

One expects that for ECO binaries a post-Newtonian (PN) description furnishes a fairly good representation of the binary's behavior in the early inspiral. 
GW signals should be well described by the point-particle binary model until tidal effects become apparent. After coalescence, it is natural to expect either an object of the same nature as the initial bodies or a BH. The GWs emitted in this regime would be related to fundamental properties of the corresponding object and/or the dynamics of the merger. However, the details of the coalescence phase, which contains precious information on the properties of the original bodies (see Sec.~\ref{obs}), are unknown. For this reason, studies on the possibility of distinguishing ECOs from standard compact objects with GWs have focused mainly on these two regimes, e.g. through perturbative modifications to the inspiral signal \cite{Sennett_2017,Maselli:2017cmm,Cardoso:2017cfl,Giudice:2017dde,Krishnendu:2017shb,Krishnendu:2018nqa,Johnson-McDaniel:2018uvs,Addazi:2018uhd,Datta:2019epe,Datta:2020gem,Asali:2020wup,Chirenti:2020bas,Pacilio:2020jza}, quasinormal modes (QNM) of ECOs \cite{Macedo:2013jja,Cardoso:2016rao,Macedo:2016wgh,Yunes:2016jcc,Chirenti:2016hzd,Maggio:2020jml} or echoes after the merger \cite{Cardoso:2016oxy,Mark:2017dnq,Maselli:2017tfq,Conklin:2017lwb,Westerweck:2017hus,Wang:2018gin,Urbano:2018nrs,Lo:2018sep,Nielsen:2018lkf,Tsang:2019zra,Maggio:2019zyv,Chen:2019hfg,Conklin:2019fcs,Micchi:2020gqy}. In order to fully exploit the potential of GW observations and ready the analysis required to discern ECOs through the full inspiral, merger and post-merger phases, we propose a toy model aiming to capture the main features of the full GW signal emitted by ECO binaries.

This paper is organized as follows. In Sec.~\ref{obs} we discuss known features of the merger of non-BH compact objects, which
we utilize to construct the toy model presented in Sec.~\ref{tm}.
We describe how the full GW signal is built in Sec.~\ref{wvf}. In Sec.~\ref{da} we assess the detectability of ECO binaries and our ability to distinguish them from standard compact binaries with current and future GW detectors. Finally, we present our conclusions in Sec.~\ref{ccl}.

\section{Coalescence of compact objects other than black holes}\label{obs}


To construct our model, we will rely on our understanding of BBHs together with BNSs and BBSs. 
In particular, the differences of the latter two systems with BBHs will inform us how to implement
in our model the particular phenomenology that general ECOs might display.
For simplicity, we focus on binaries consisting of (nonspinning) identical objects (same nature and same mass) on quasicircular orbits. Here and throughout this paper, we denote by $m_0$ and $r_0$ the mass and radius of the bodies, and their compactness in isolation by $C_0=Gm_0/r_0c^2$.

\begin{figure}[th!]
{\centering \includegraphics[scale=0.095]{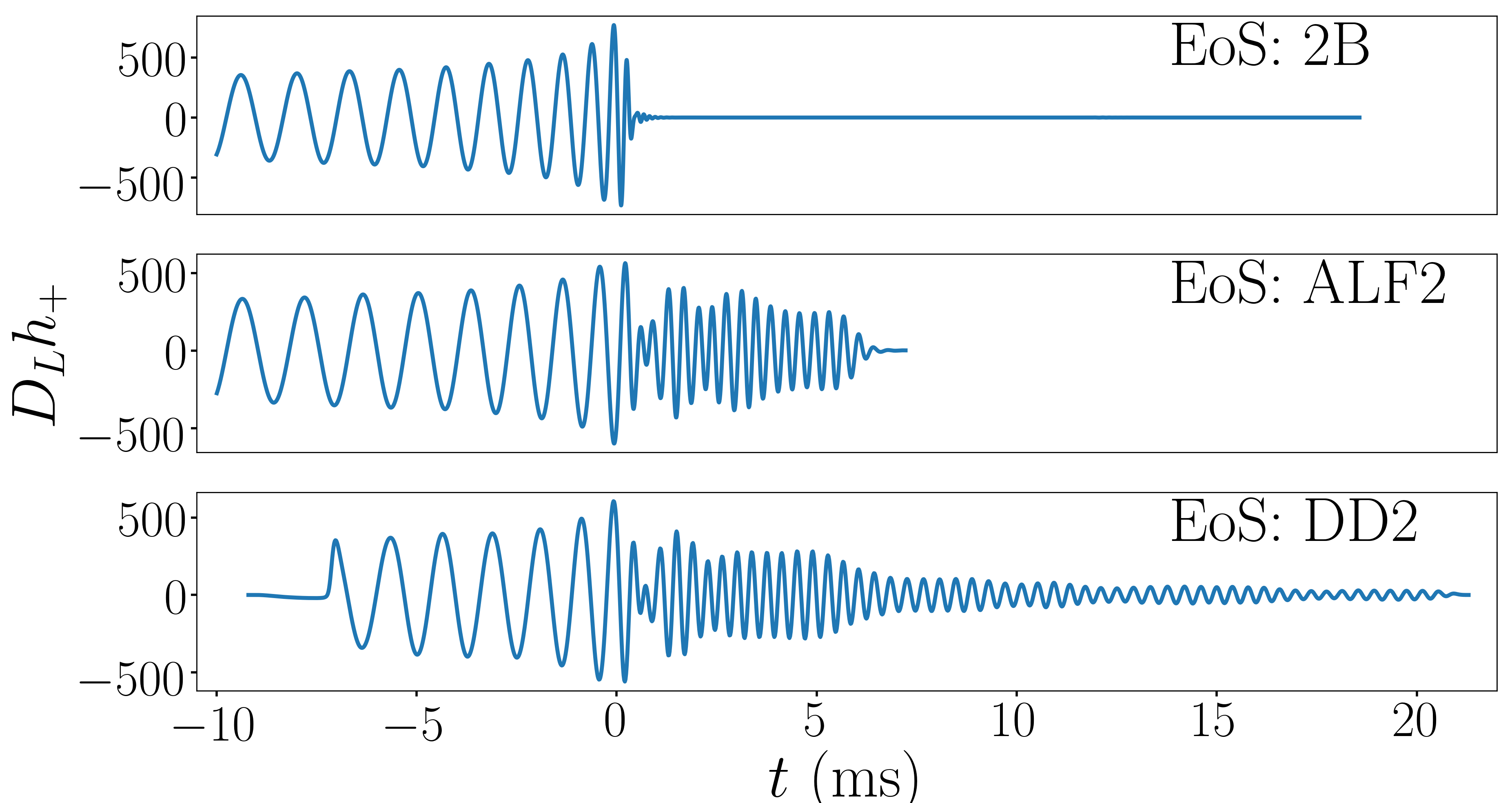}
}
\caption{Numerical simulations of BNS waveforms for different EoSs. For the 2B EoS, the system promptly collapses to a BH after contact. For the ALF2 and the DD2 EoS, a hypermassive NS is formed following contact. In the former case the hypermassive NS ends up collapsing to a BH after $\simeq 7$~ms, whereas in the latter case it does not collapse for the duration of the simulation, seemingly relaxing to a stable NS.}\label{wvf_num}   
 \end{figure}

During the inspiral phase of BNSs and BBSs, tidal effects lead to a correction in the GW phase at 5PN order relative to BBHs \cite{1984grg..conf...89D,Flanagan_2008,Wade:2014vqa} (in GR). This correction is proportional to the dimensionless tidal polarizability: $\Lambda=\frac{2}{3}k_2 C_0^{-5}$ \cite{Flanagan_2008,Wade:2014vqa}, $k_2$ being the tidal Love number. The compactness and the tidal Love number are determined by the equation of state (EoS). NSs typically have $C_0$ in the range $0.14-0.2$ (see Fig.~7 of \cite{_zel_2016}). To date, proposed models for nonspinning BSs give similar values, but can reach higher compactness (so far, up to $0.3$ ~\cite{Friedberg:1986tq,Kesden:2004qx}). As for $k_2$, it is typically of the order of $0.1$ for NSs and currently available models of BSs \cite{Damour:2009vw,Binnington_2009,Hinderer_2008,Cardoso:2017cfl,Sennett_2017}. 

The PN description fails when the stars come in contact. This approximately takes place at the ``contact frequency''
\begin{align}
 f_c=\frac{c^3}{2G}\frac{C_0^{3/2}}{m_0}, \label{eq_fc}
\end{align}
which depends on the EoS through the compactness. For low enough values of $C_0$ ($\lesssim 0.29$), this frequency is lower than
that of the innermost stable circular orbit for a BBH of the same mass. For this reason, we will employ the terminology ``post-contact'' 
rather than merger/post-merger whenever appropriate. For ultracompact exotic objects, $C_0\sim 0.5$ and the objects reach the innermost stable circular orbit before touching. We thus expect the merger to proceed in a fashion more similar to BBHs, with no post-contact stage. However, no numerical simulations in this regime are available so far to confirm this expectation. Moreover, it is possible that such objects do not exist in nature as they might be nonlinearly unstable \cite{Cardoso:2014sna}.  
After contact, a rather complex behavior is displayed by BNSs and BBSs (and thus potentially also by ECO binaries), while the system's evolution
is comparatively simpler in the case of BBHs. This behavior is only explorable through numerical simulations. 

After the stars touch, the least bound material might be disrupted (and a portion of it might even be ejected), whereas the remaining part gives rise to an envelope containing the inner cores of the original bodies \cite{Shibata_2000,Palenzuela:2017kcg}. Gravitational interaction acts to bring the cores together, but internal restoring forces (whose exact nature depends on the objects) and angular momentum oppose it. As a result, 
collapse to a BH can be prevented, forming instead a hypermassive star \cite{Baumgarte:1999cq,Palenzuela:2017kcg}. As these effects compete, the cores oscillate \cite{Shibata_2000,Palenzuela:2017kcg}. Eventually, collapse to a BH or formation of a stable star ensues. The exact outcome is determined by the total mass of the system, the EoS and
any further relevant physics at play (e.g.~microphysics, electromagnetic effects and gravitational cooling)\cite{Bernuzzi:2020tgt,Palenzuela:2006wp,Lehner:2014asa,Bezares:2018qwa}.

 \begin{figure}[th!]
{\centering \includegraphics[scale=0.095]{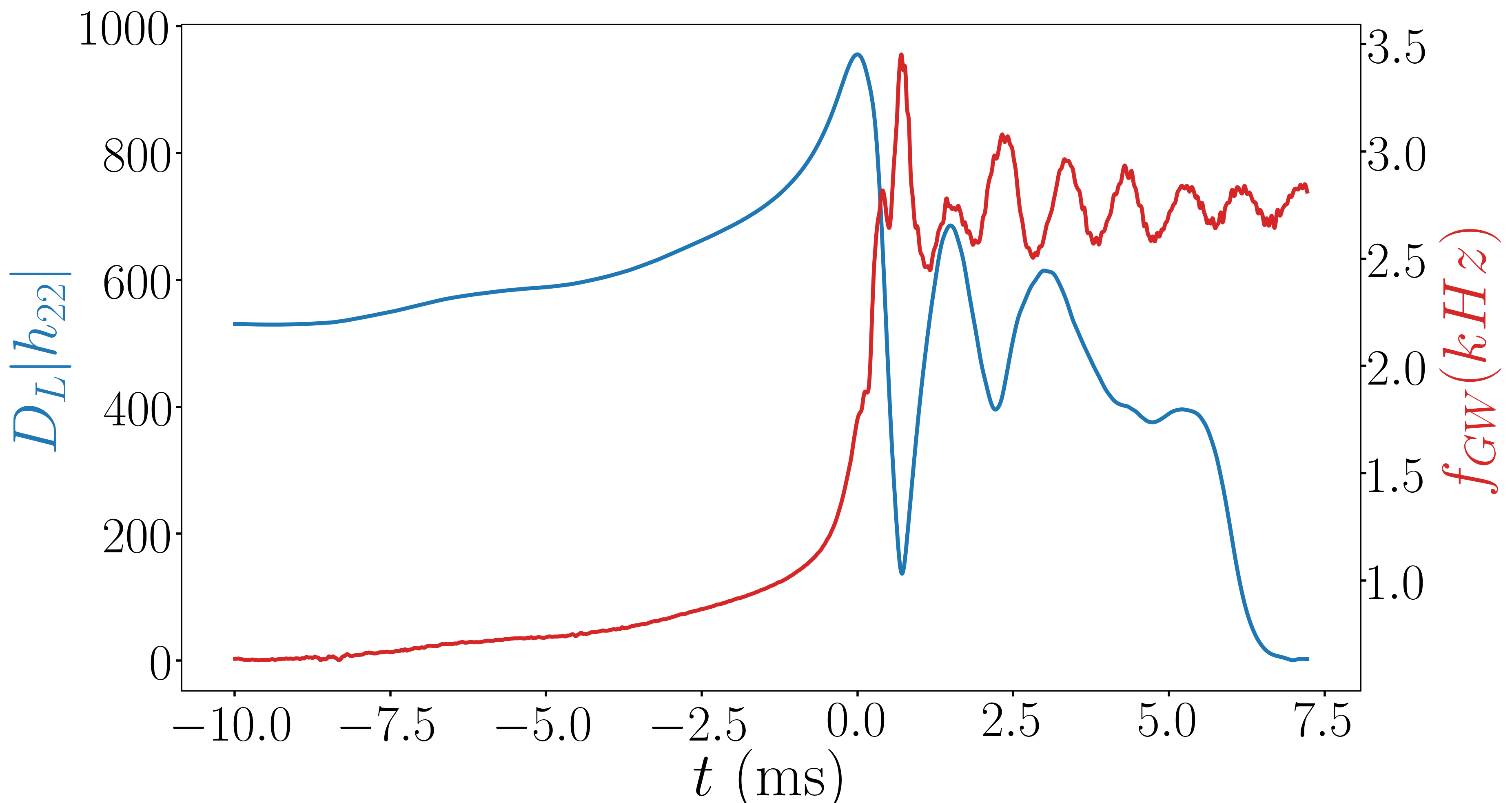}
}
\caption{Amplitude-frequency decomposition of the GW signal emitted by a BNS following the ALF2 EoS. The signal is characterized by the oscillations in amplitude (blue) and frequency (red) in the post-contact stage.}\label{wvf_decomp_num}   
 \end{figure}

These different scenarios leave distinct imprints on the waveform, as can be seen in Fig.~\ref{wvf_num}. 
There, we display the GW signal produced by the merger of two NSs with initial masses $m_0=1.35 \ {M_{\odot}}$, for different choices of the EoS: 2B \cite{Read:2008iy}, ALF2 \cite{Read:2008iy} and DD2 \cite{Typel:2009sy}. The merger happens at $t=0$. The waveforms that we show were taken from the \mbox{\texttt{CoRe} database~\footnote{http://www.computational-relativity.org/.} \cite{Dietrich:2018phi}}. The system in the upper panel collapses immediately to a BH following contact, whereas the other two form a hypermassive NS. The system in the middle panel collapses to a BH after $\sim 7 \ {\rm ms}$, and the one in the lower panel does not collapse for the simulation duration, seemingly forming a stable NS. In the last two cases, the difference with a BBH signal is visible with naked eye. Fig.~\ref{wvf_decomp_num} shows the decomposition of the signal obtained with the ALF2 EoS (middle panel of Fig.~\ref{wvf_num}) into amplitude and frequency. We note that the post-contact signal is characterized by oscillations in the GW amplitude and frequency.

 Remarkably, for some models of BSs, if the outcome is a BS, angular momentum is entirely radiated immediately after contact, with very little mass ejected~\cite{Palenzuela:2017kcg,Bezares:2017mzk,Sanchis-Gual:2019ljs,DiGiovanni:2020ror}. The hypermassive BS radiates GWs primarily at the fundamental
quasinormal frequency of the star~\cite{Palenzuela:2017kcg}. On the other hand, for BNSs the main frequencies in the post-contact stage are determined by the dynamics of the binary \cite{Bauswein_2012_1,Bauswein_2012_2,Bauswein:2015yca,Lehner_2016}. 



With these observations in mind, we propose that the coalescence of ECO binaries should consist of 
(i) an inspiral phase with tidal effects (ii) a post-contact evolution with three 
possible scenarios:
\begin{itemize}
 \item prompt collapse to a BH,
 \item formation of a hypermassive ECO that ultimately collapses to a BH,
 \item formation of a hypermassive ECO that settles into a stable remnant of the same nature as the original bodies.  
\end{itemize}
Moreover, we expect the amount of angular momentum retained in the post-contact evolution to depend on the nature of the original bodies. 
We introduce a parameter $ 0 \leq \kappa \leq 1$ that represents the fraction of angular momentum retained at the onset of the post-contact phase. This parameter depends on the nature of the compact objects, and it is related to the ability of a given compact object to sustain rotation. For instance, it would be $\sim 1$ ($\sim 0$) in case a stable remnant forms from the coalescence of a BNS (some models of BBS).
This generic classification allows for a model agnostic phenomenological description of the coalescence of ECO binaries, irrespective of the exact nature of the involved bodies. 
In this work, we focus on the last two scenarios, and consider only extremal values of $\kappa$: we take $\kappa=1$ for the scenario where a BH is formed as the result of the collapse of a hypermassive ECO, and $\kappa=1$ or $\kappa=0$ for the scenario where a stable remnant is formed. Thus, we consider three types of behavior in the post-contact stage:
\begin{itemize}
 \item \emph{RBH}: \emph{rotating} systems that collapse to a \emph{BH}
 \item \emph{RS}: \emph{rotating} systems that form a \emph{stable} remnant
 \item \emph{NRS}: \emph{non-rotating} systems that form a \emph{stable} remnant
\end{itemize}
We focus on bodies with compactness in the range $0.14 \lesssim  C_0 \lesssim 0.2$, which contains the typical values of many models of stable compact objects other than BHs. For high values of $C_0$, the toy model that we present in the next section predicts a prompt collapse to a BH, a scenario on which we do not focus. This is in agreement with the numerical simulations available so far, which suggest that for $C_0$ higher than $\sim 0.18$, the coalescence of BBSs leads to a BH \cite{Palenzuela:2017kcg,Bezares:2017mzk,Bezares:2018qwa}. However, one should keep in mind that for EoSs that have not been explored yet, or for different setups, it might be possible to form more compact remnants. In particular, stable BSs can reach compactnesses of $\sim 0.3$ \cite{Friedberg:1986tq,Kesden:2004qx}, suggesting such endstates might be viable. (Objects with higher compactness, however, might not be stable \cite{Cardoso:2014sna}, in which case they
would not be a possible coalescence  end product.) Our framework could be modified to account for these cases, e.g. by attaching a rotating bar instead of the toy model presented in this work.




\section{Toy model} \label{tm}
\begin{figure}[th!]
{\centering \includegraphics[scale=0.2]{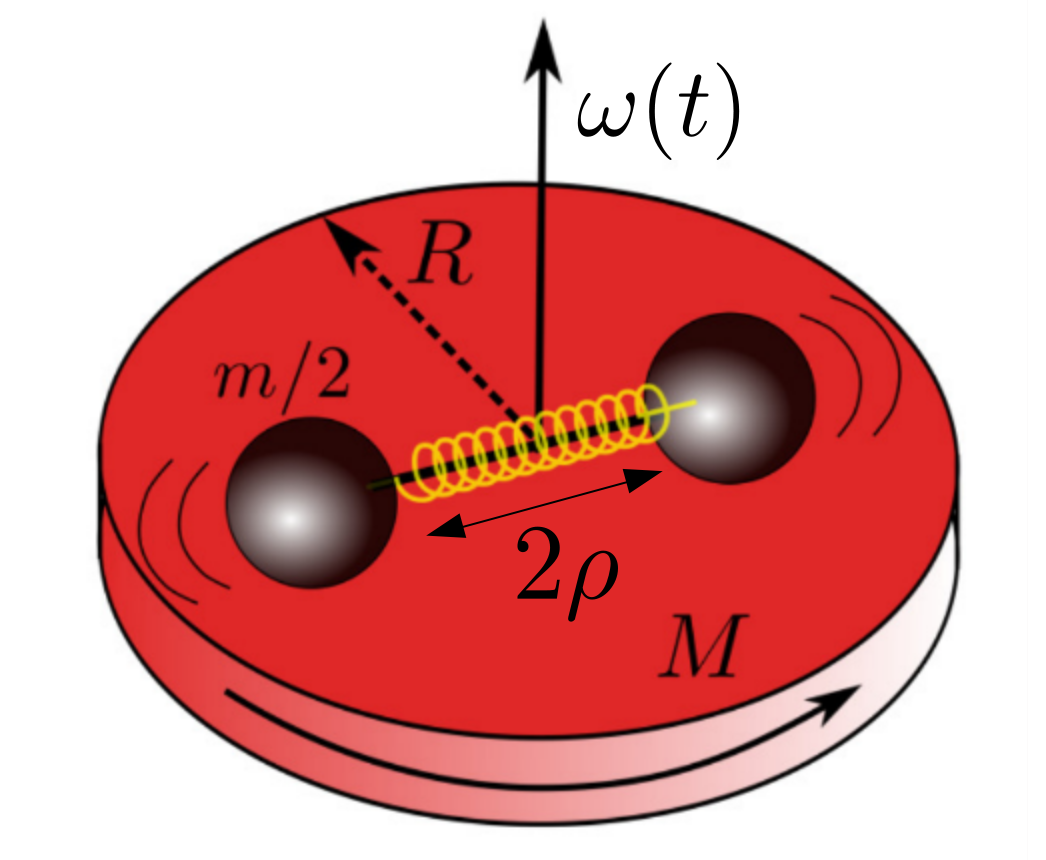}
}
\caption{Illustration of the setup we use to model the dynamics of the post-contact stage, adapted from \cite{Takami:2014tva}. The hypermassive ECO is treated as a disk containing two point particles that interact gravitationally, but also through an effective spring. The latter mimics the effect of the restoring forces, making the cores bounce.}\label{fig_tm}   
 \end{figure}
 
Our starting point is the toy model introduced in \cite{Takami:2014tva} to describe the post-contact dynamics of BNSs. We model the inner cores of the ECOs by point particles interacting gravitationally, but also through an effective spring. The latter mimics the effect of the restoring forces, making the cores bounce. The disrupted material is modeled by a disk containing the two point particles, see Fig.~\ref{fig_tm}. For \emph{RBH} and \emph{RS} systems the disk and the cores corotate, for \emph{NRS} systems neither the disk nor the cores rotate.

The toy model is characterized by four free parameters: the radius of the disk ($R$), the mass of the cores ($m/2$), the spring constant ($k$) and its length at rest $2 \rho_0$. We assume mass conservation (i.e. no ejection of mass following contact) and that the two point particles have the same mass $m/2$. Therefore, the mass of the disk is $M=2m_0-m$, and its radius satisfies $r_0 \leq R \leq 2 r_0$. We define the characteristic frequency of the spring as $\omega_0=\sqrt{k/m}$. These free parameters should be related to the EoS of the ECOs in the binary, e.g.~higher values of $k$ correspond to stiffer EoSs (see \cite{Lucca:2020iax} for a discussion in the case of BNSs). Within this toy model, the three types of post-contact evolution (\emph{RBH}, \emph{RS} and \emph{NRS}) are distinguished by the choice of free parameters and the prescription for the angular momentum. 

The dynamics is governed by four variables: the distance of each core to the center ($\rho$); 
the orbital angle ($\phi$), out of which we define the orbital angular velocity ($\omega=\dot{\phi}$); the total energy ($E$); and the total angular momentum ($J$). 
Quantities labeled by $c$ refer to the time of first contact, and those labeled by $i$ refer to the beginning of the toy model. We assume that, due to the finite shock-propagation speed, after the two objects touch, only their inner parts are compressed, whereas the outer layers keep their original sizes. We can therefore write $\rho_i=R-r_0$. We assume energy conservation and assume $J_i=\kappa J_c$, where $\kappa$ is 0 for \emph{NRS} systems and 1 otherwise. The initial orbital phase plays no role in the dynamics, and will be used later to match the post-contact stage to the inspiral.  

\begin{figure}[th!]
 \centering
   \includegraphics[scale=0.095]{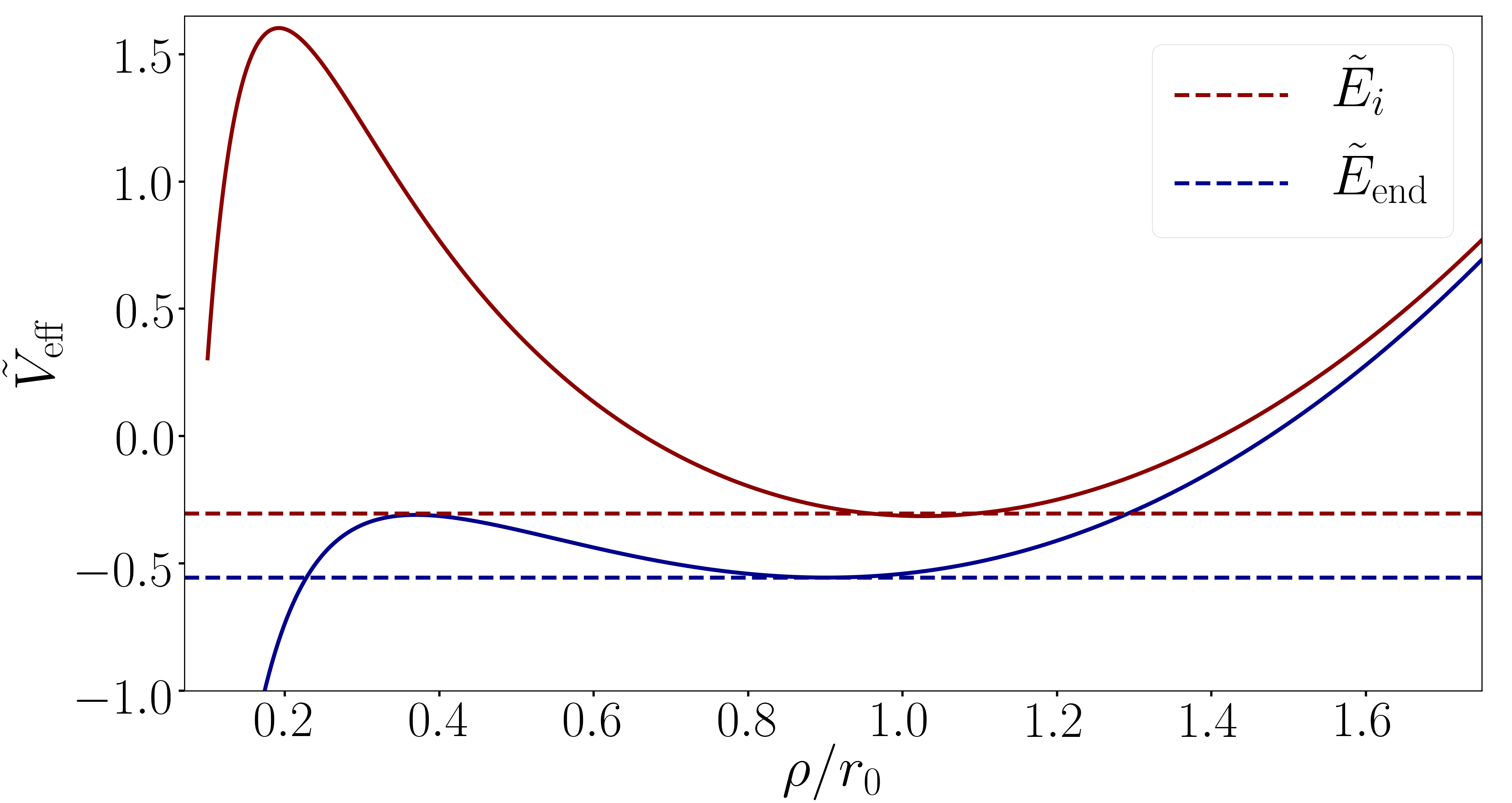}
   \caption{Evolution of the effective potential and the energy for an \emph{RS} system. As the evolution proceeds, the energy reaches the minimum of the potential, leading to a ``circularization'' of the orbit.}
   \label{ev_pot}
 \end{figure}

We map the dynamics to that of an effective particle of mass given by the system's reduced mass, evolving in a potential well $V_{{\rm eff}}=V_{{\rm centrifugal}}+V_{{\rm spring}}+V_{{\rm gravitational}}$. The gravitational term was not accounted for in \cite{Takami:2014tva}, and its expression, together with the centrifugal force and the spring terms, is given in App.~\ref{app:tm}. The addition of the gravitational term allows for a wider variety of post-contact behaviors, and helps provide a natural description of collapse to a BH. Moreover, the treatment of \cite{Takami:2014tva} can be seen as a particular case of ours, in the limit of high values of $k$ (stiffer objects), where the spring term dominates over the gravitational one. The shape of the effective potential is determined by the free parameters of the model. In Fig.~\ref{ev_pot} we display by dark-red lines the effective potential and the energy (in rescaled dimensionless units) at the start of the post-contact stage, for a given choice of the free parameters (corresponding to an \emph{RS} system). The effective particle is trapped between an inner and an outer turning point, reproducing the oscillatory behavior described in Sec.~\ref{obs}. As the system evolves, energy and angular momentum are dissipated through radiation of GWs. Due to the dependence of the centrifugal potential on the angular momentum, the effective potential changes during the evolution, as illustrated by the solid dark-blue line in Fig.~\ref{ev_pot}.

The evolution of the effective particle is governed by
\begin{align}
& \dot{\varphi}=\omega  = \frac{J}{I}=\frac{J}{\frac{MR^2}{2}+m\rho^2}, \label{eq_omega_gw} \\ 
 & \dot{\rho}  = \pm  \sqrt{\frac{2}{m}} \sqrt{E-V_{\rm eff}(\rho)},  \label{eq_rho_gw} \\
 & \dot{E}  =-P_{\rm GW}, \label{eq_e_gw} \\
 & \dot{J}  =-\dot{J}_{\rm GW} \label{eq_j_gw}.
\end{align}
We compute the energy and angular momentum carried away by GWs consistently, using the adiabatic approximation and averaging over an entire period of the radial motion. 
Details of the computation are given in App.~\ref{app:gw}. 
We integrate the adiabatic equations of motion, Eqs.~\eqref{eq_omega_gw}, \eqref{eq_rho_gw}, \eqref{eq_e_gw} and \eqref{eq_j_gw}, over successive periods, see Apps.~\ref{app:ecc} and \ref{app:circ} for details. The evolution proceeds differently for each type of ECO binary:
%

\begin{itemize}
 \item For \emph{NRS} systems, one has $J=0$ and thus the potential is fixed. As the evolution proceeds, the effective particle loses energy and sinks to 
the bottom of the potential. Physically, this corresponds to the formation of a stable object of the same nature as the binary's ECOs, with zero angular momentum. We fix the parameters of the toy model to match the frequency of the radial oscillations to numerical simulations of BBSs \cite{Palenzuela:2017kcg} (in principle we could also use QNMs of different ECO models, although numerical simulations are needed in order to check that these are the relevant frequencies in this regime, as was explicitly done in \cite{Palenzuela:2017kcg}.)

\item For \emph{RS} systems, the effective particle sinks to the bottom of the potential (see Fig.~\ref{ev_pot}) and the orbit ``circularizes''. The system keeps emitting GWs as it settles into an equilibrium state, corresponding to the formation of a stable object of the same nature as the original ECOs.

\item For \emph{RBH} systems, the energy eventually becomes larger than the potential height, leading to a ``plunge'' and subsequent formation of a BH. Following the hoop conjecture \cite{hoop_conj}, we assume that a BH is formed when the compactness of the system becomes larger than $0.5$ (the minimum compactness of a BH). The compactness of the system is defined as
 \begin{equation}
 C_{\rho}=\frac{GM_{\rho}}{\rho c^2}, \label{comp}
\end{equation}
where $M_{\rho}=m+M(\rho/R)^2$ is the total mass within the radius $\rho$. 
The BH thus formed has mass $M_{f}=M_{\rho}$ and dimensionless spin parameter $a_{f}=cJ_{\rho}/GM_{\rho}^2$, where $J_{\rho}=(m+M(\rho/R)^2) \rho^2 \omega$ is the angular momentum of the collapsing matter out of which the BH forms.
\end{itemize}

We display examples of the post-contact dynamics of an \emph{RBH} and an \emph{RS} system in App.~\ref{app:dyn}.
We found the results of this toy model to be more sensitive to the compactness rather than to the exact values of the free parameters. In the remainder of this paper, we present results obtained with choices of the free parameters that we consider to be representative of each type of ECO binary. 

\section{Waveform}\label{wvf}

\subsection{Time domain}

We focus on the dominant 22-mode of the GW signal emitted by an ECO binary and match smoothly its amplitude and phase across the different stages of the evolution. 
We recall that for a GW signal emitted along the direction of the orbital angular momentum, one has $h_{22}=\sqrt{\frac{4\pi}{5}}(h_{+}-ih_{\times})$. Defining $h_{22}=A(t)e^{-i\Psi(t)}$, the instantaneous GW frequency is given by $\hat{f}=\frac{1}{2\pi}\frac{{\rm d} \Psi}{{\rm d} t}$.

\subsubsection{Inspiral}\label{wvf_inspiral}
 
For the inspiral, we use the effective-one-body waveform SEOBNRv4T, an extension of SEOBNRv4 \cite{Boh__2017} accounting for tidal effects\footnote{This approximant is available in the LIGO Algorithm Library \cite{lalsuite}.}, and take $k_2=0.1$. We stop the inspiral waveform at contact, i.e.~when $\hat{f}=2f_c$. We assume that the formation of the hypermassive star happens over a time interval $\Delta t=1/2f_c$, which corresponds to half a period after contact. For $m_0=1.35 \ M_{\odot}$, this gives $\Delta t \simeq 0.7\ {\rm ms}$, in good agreement with numerical simulations of BNSs. The inspiral is matched to the post-contact waveform over this interval, using cubic functions to ensure smoothness of the amplitude and the instantaneous GW frequency.

\subsubsection{Post-contact}

During this stage the waveform is computed using the quadrupole formula, see App.~\ref{app:gw}. It yields
\begin{align} 
h_{+}&=\frac{2}{D_L}\frac{G}{c^4}(\gamma_2 \cos(2\phi)-\gamma_1 \sin(2\phi)), \label{eq_hp} \\
h_{\times}&=\frac{2}{D_L}\frac{G}{c^4}(\gamma_1 \cos(2\phi)+\gamma_2 \sin(2\phi)) \label{eq_hc},
\end{align}
where
\begin{align}
 \gamma_1=&\frac{2m \rho \dot{\rho} \omega (MR^2+ m \rho^2) }{\frac{MR^2}{2}+ m \rho^2}, \label{def_f1} \\
 \gamma_2=&   m \left ( \dot{\rho}^2 -\rho^2 \omega^2 - 4 \omega_0^2 \rho (\rho-\rho_0) -\frac{Gm}{8\rho}+\frac{GM\rho}{R^2} \right ) \label{def_f2}. 
\end{align}
The phase and amplitude of the 22-mode are given by
\begin{align}
 A(t)&=\sqrt{\frac{16 \pi}{5}}\frac{G}{D_Lc^4} \sqrt{\gamma_1^2+\gamma_2^2}, \label{amp22} \\
 \Psi(t)&=2\phi-\arctan(\gamma_1/\gamma_2). \label{ph22}
\end{align}
The initial orbital phase ($\phi_i$) is fixed by matching to the inspiral.
For \emph{NRS} systems one has $\gamma_1=0$ and $\phi=\phi_i$; therefore, with our definition, the instantaneous frequency would vanish in the post-contact phase. In this particular case, we define it instead as the frequency of radial oscillations. 


\subsubsection{Ringdown}

For \emph{RBH} systems, we model the ringdown signal of the final BH using the model presented in \cite{Bohe:2016gbl,Damour:2014yha},
 \begin{equation}
  h^{\textrm{RD}}_{22}(t) = \eta\, \tilde{A}_{22}(t)\,e^{i \tilde{\phi}_{22}(t)}\,e^{-i \sigma_{220} (t-t_{\textrm{match}}^{22})},
 \end{equation}
 where $t_{\textrm{match}}^{22}$ is the matching time and $\sigma_{220}=-\sigma^{I}+i\sigma^{R}$ is the least damped QNM frequency of a perturbed Kerr BH, computed from by \cite{Berti:2009kk}. The BH mass and spin are computed as described below Eq.~\eqref{comp}. The functions $\tilde{A}_{22}(t)$ and $\tilde{\phi}_{22}(t)$ are defined by
 \begin{equation}
  \tilde{A}_{22}(t) \equiv c_1^c \tanh\left[c_1^f (t - t_{\textrm{match}}^{22}) + c_2^f\right] + c_2^c,
 \end{equation}
and 
 \begin{equation}
  \tilde{\phi}_{22}(t) = \phi_1 - d_1^c \log\left[\frac{1+d_2^f e^{-d_1^f (t-t_{\textrm{match}}^{22})}}{1+d_2^f}\right].
 \end{equation}
Coefficients with superscripts $f$ are calibrated against numerical simulations (the expressions are given in \cite{Bohe:2016gbl,Damour:2014yha}). Those with superscripts $c$ are used to ensure continuity of the amplitude, phase and their first derivatives at the matching time, while $\phi_1$ is the phase at the start of the ringdown.
With these matching prescriptions, the phase and the amplitude are $\mathcal{C}^1$ functions.

Our model could also incorporate different behaviors for the final stage, such as resonant modes of excited stars for \emph{RS} systems, but we do not explore this possibility in this work.

\subsubsection{Full waveform}
\begin{figure}[th!]
{\centering \includegraphics[scale=0.09]{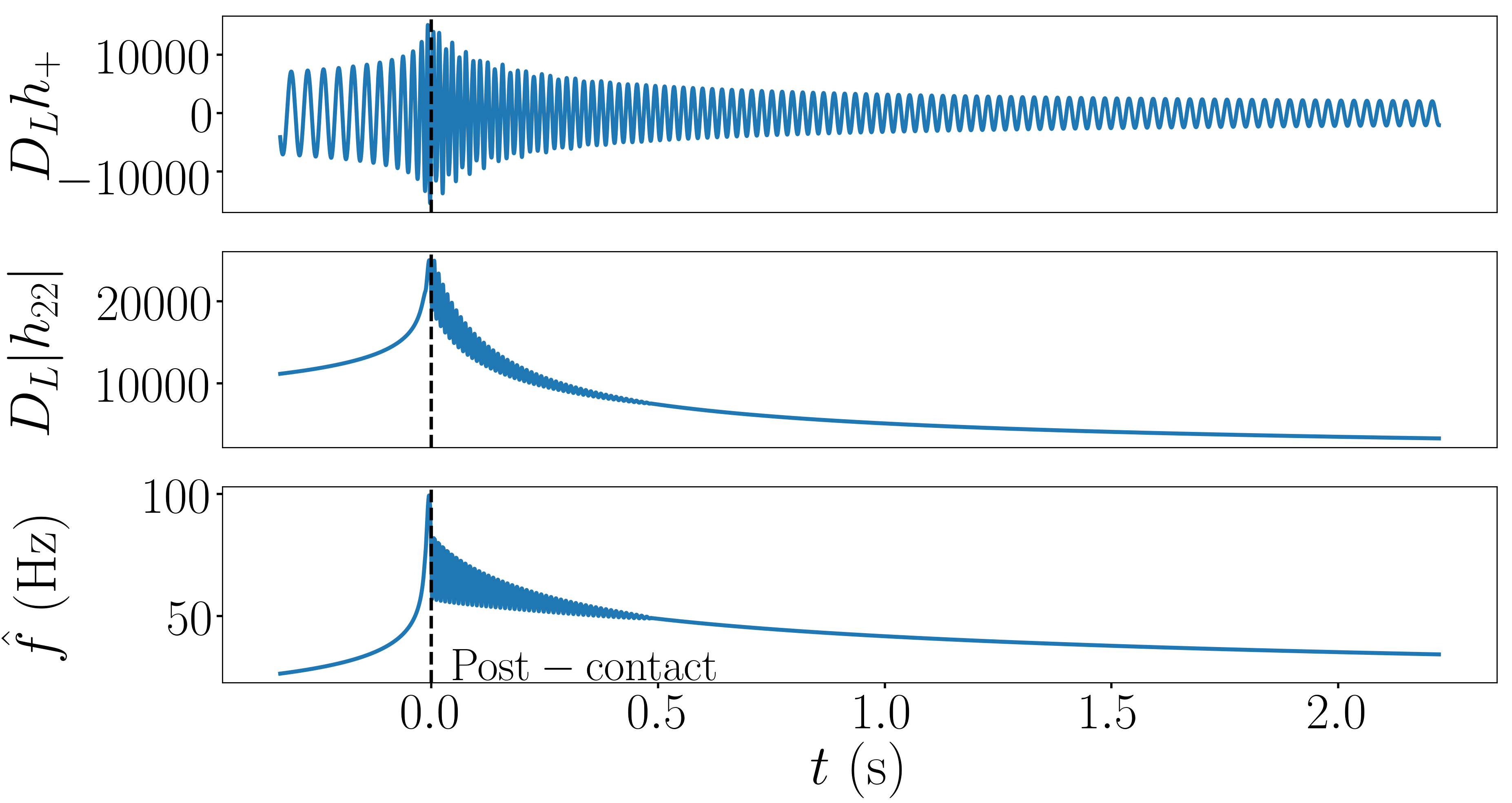}
}
\caption{Waveform for an \emph{RS} system with $m_0=30 \ M_{\odot}$ and $C_0=0.17$. We show the real part of the 22-mode and its decomposition into amplitude and frequency. The black dashed lines indicate the start of the post-contact stage.}\label{wvf_td_ns}   
 \end{figure}
\begin{figure}[th!]
{\centering \includegraphics[scale=0.09]{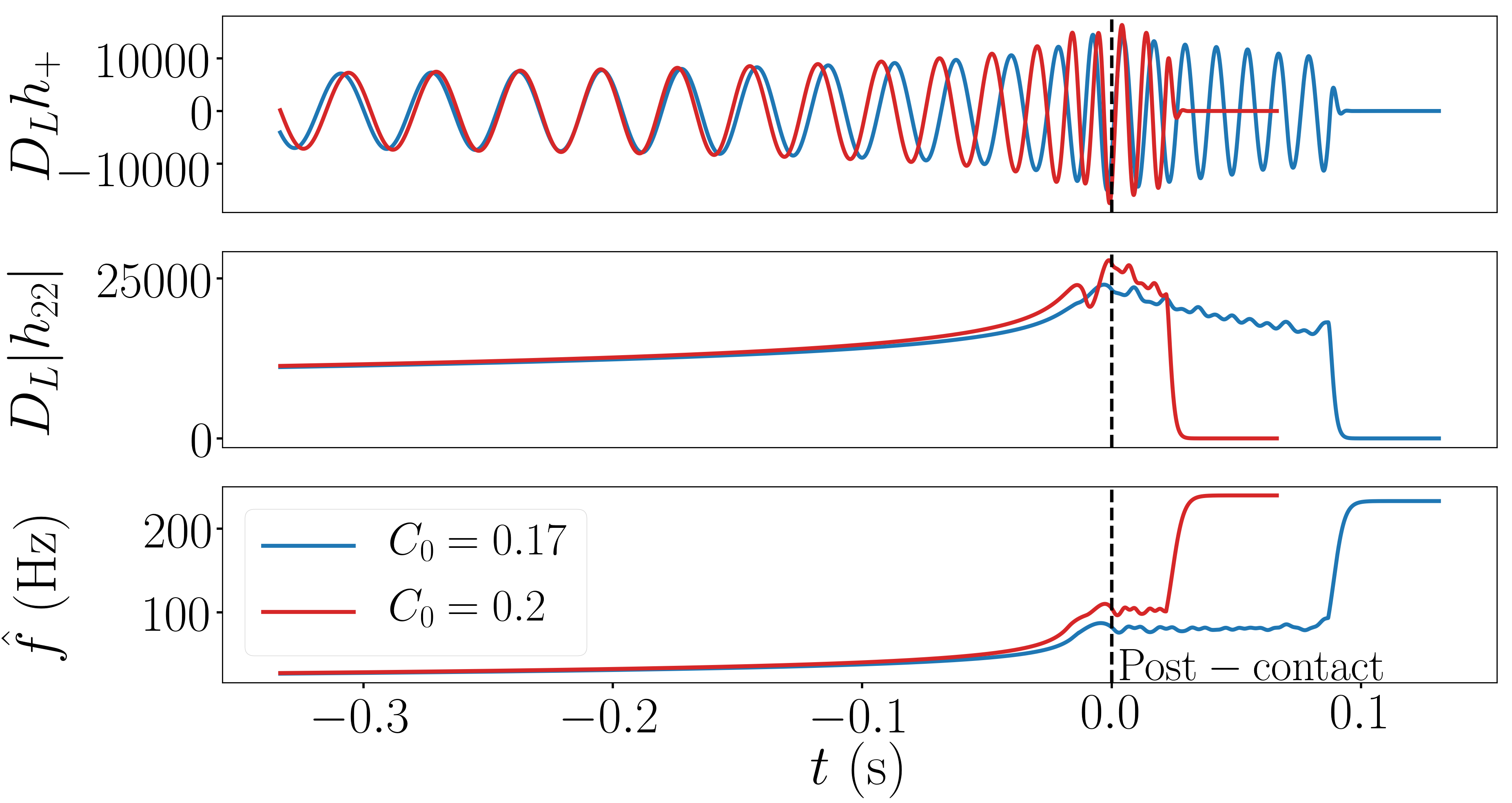}
}
\caption{The same as Fig.~\ref{wvf_td_ns}, but for an \emph{RBH} system with $m_0=30 \ M_{\odot}$ and $C_0=0.17$ ($C_0=0.2$) in blue (red). After contact, the collapse to a BH happens faster for more compact systems.}\label{wvf_td_bh}   
 \end{figure}

\begin{figure}[th!]{
    \centering \includegraphics[scale=0.09]{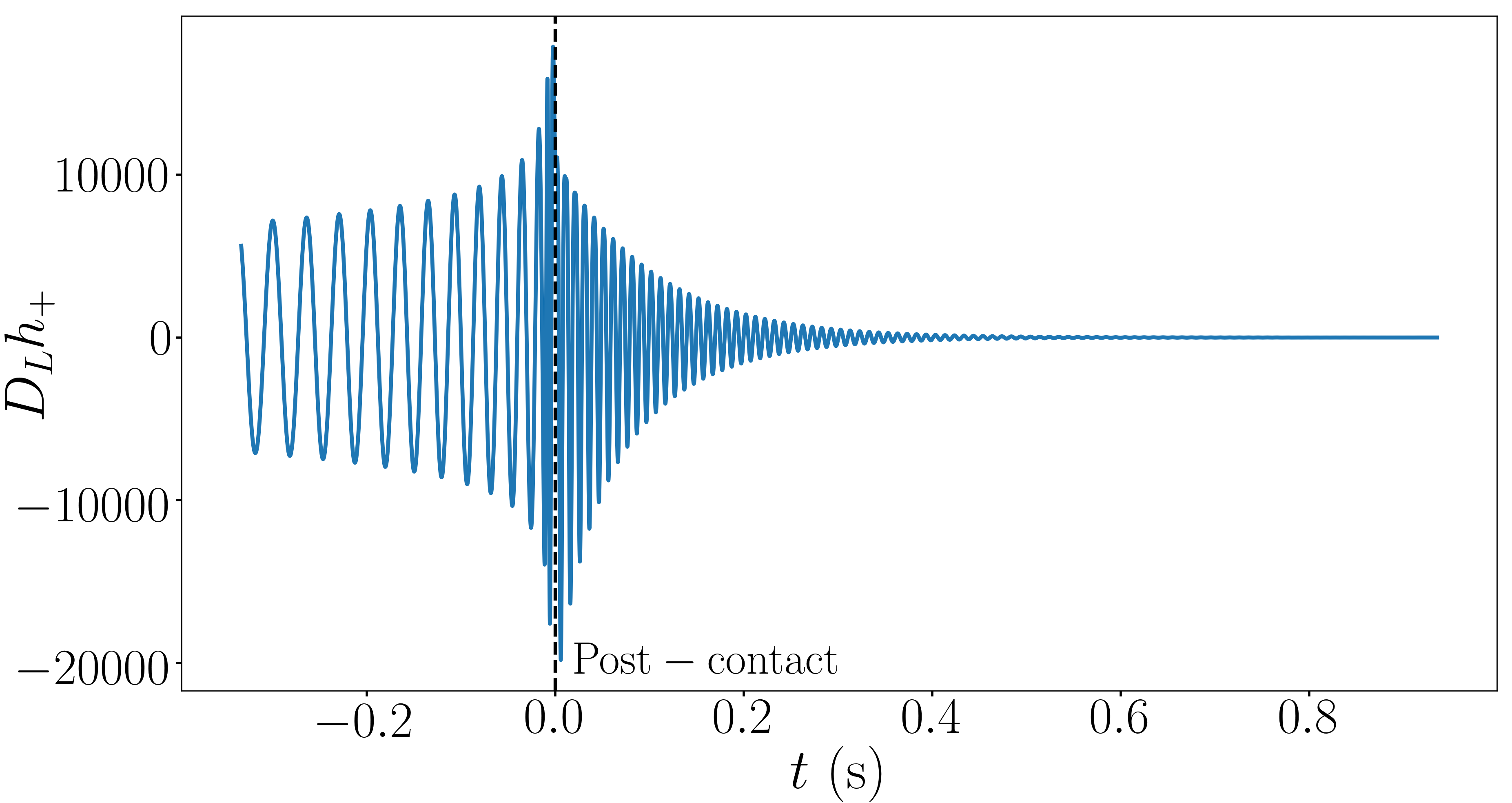}
    }
    \caption{Real part of the 22-mode for an \emph{NRS} system with $m_0=30 \ M_{\odot}$ and $C_0=0.17$.}\label{wvf_td_bs} 
 \end{figure}

 
 In Figs. \ref{wvf_td_ns} and \ref{wvf_td_bh} we display the real part of the 22-mode, together with its decomposition in amplitude and frequency for an \emph{RS} and an \emph{RBH} system with $m_0=30 \ M_{\odot}$ and $C_0=0.17$. For the sake of clarity, we display only the last instants of the inspiral. The start of the post-contact signal is indicated with black dashed lines. Our model reproduces qualitatively the oscillatory behavior of the GW amplitude and frequency in the post-contact stage. For the \emph{RBH} system, $\hat{f}$ increases up to the ringdown frequency, whereas for the \emph{RS} system it tends to zero as the system settles into a stable ECO. 
For completeness, in Fig.~\ref{wvf_td_bs} we show the real part of the 22-mode for an \emph{NRS} system with the same mass and compactness.

In Fig.~\ref{wvf_td_bh} we overplot in red the evolution for a more compact system ($C_0=0.2$). This comparison has to be performed carefully: if we were to align the waveforms at the same reference frequency, the signal of the less compact binary would be shorter, because less compact objects touch earlier. Here, we aligned the waveforms at their respective contact frequencies to highlight the post-contact evolution (the apparent alignment $\sim 0.25 \ {\rm s}$ before the merger is coincidental). After contact, the collapse to a BH happens faster for more compact objects. Moreover, more compact binaries emit GWs in the post-contact stage at higher frequencies, as expected since their contact frequency is higher.

 \subsection{Frequency domain} \label{wvf_fd}

 \begin{figure}[th!]
\subfigure[In the post-contact stage, \emph{NRS} systems emit almost exclusively at the frequency of radial oscillations and the corresponding higher harmonics.]{
    \centering \includegraphics[scale=0.09]{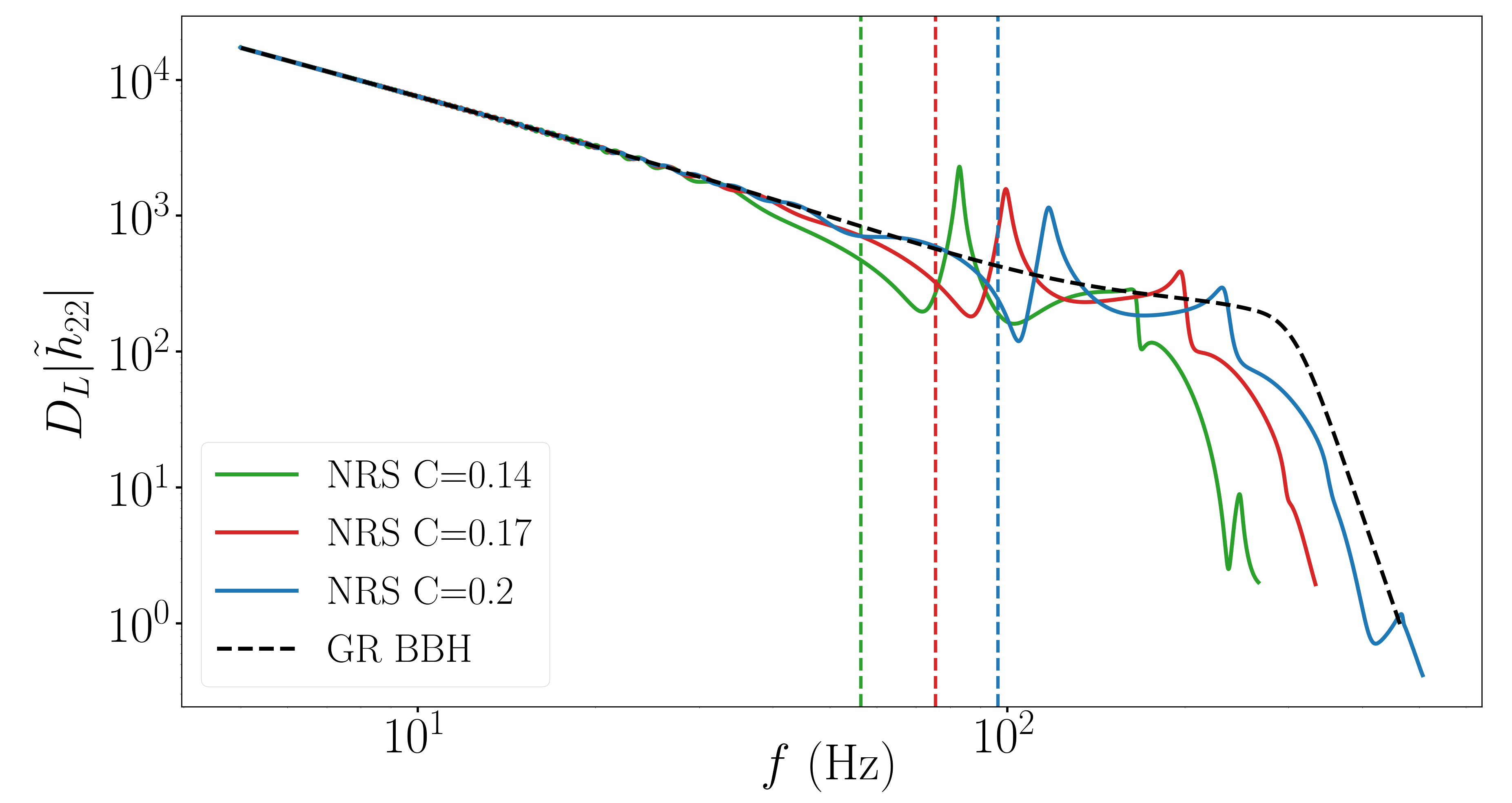}
    }
\hspace{2em}
\subfigure[For \emph{RBH} systems, we observe a main peak at the frequency around which $\hat{f}$ oscillates in Fig. \ref{wvf_td_bh}. There is also considerable power emitted at lower and higher frequencies. Thus, the inspiral and post-contact signals ``interfere'', leading to wiggles around $2f_c$. ]{
    \centering \includegraphics[scale=0.09]{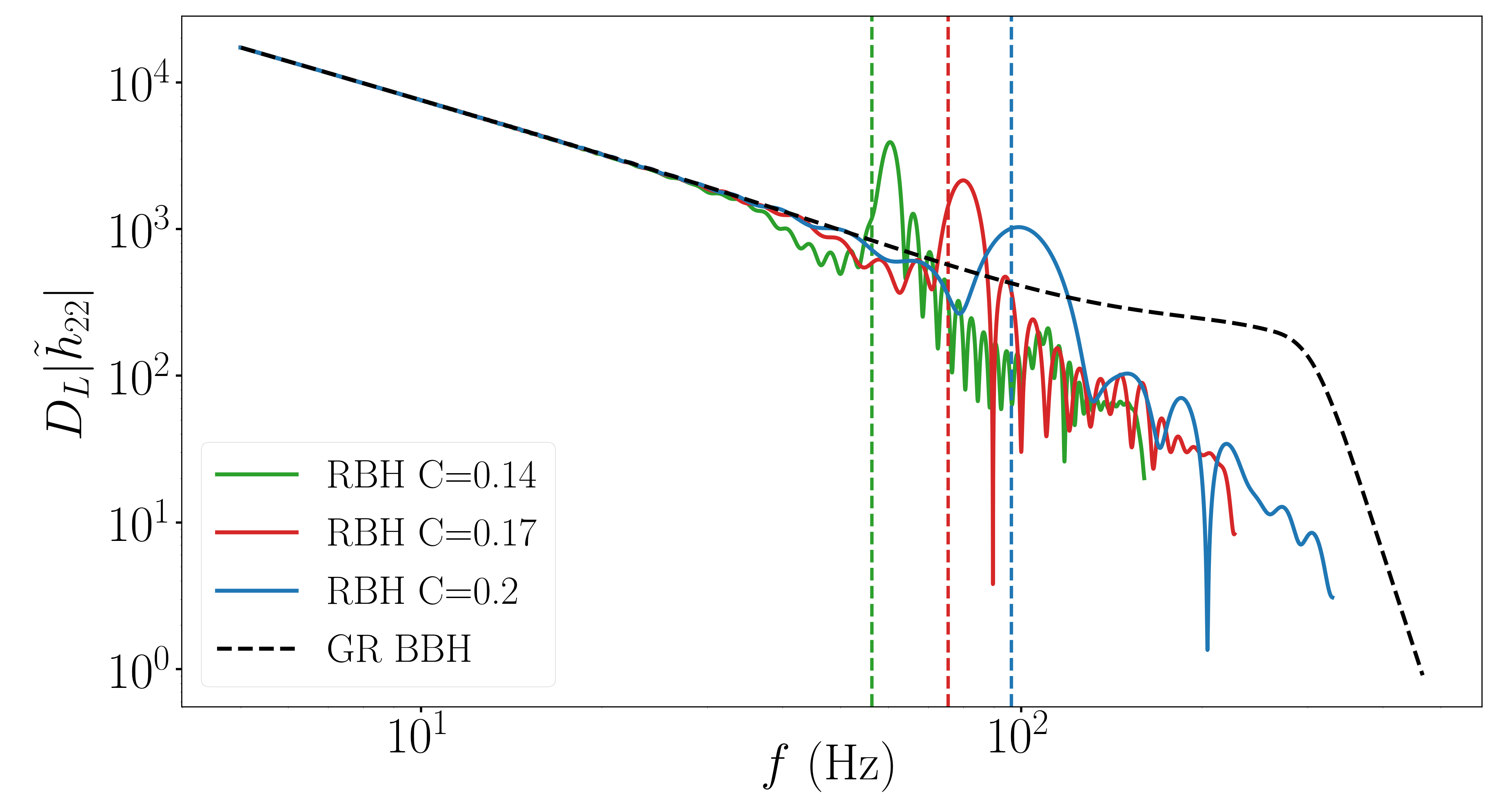}
   }
   \subfigure[For \emph{RS} systems, $\hat{f}$ tends to zero (see Fig.~\ref{wvf_td_ns}), and the peak observed in \emph{RBH} systems is therefore spread over lower frequencies, leading to significant ``interference'' with the inspiral signal. The repetition of an oscillatory pattern can be observed at higher frequencies.]{
    \centering \includegraphics[scale=0.09]{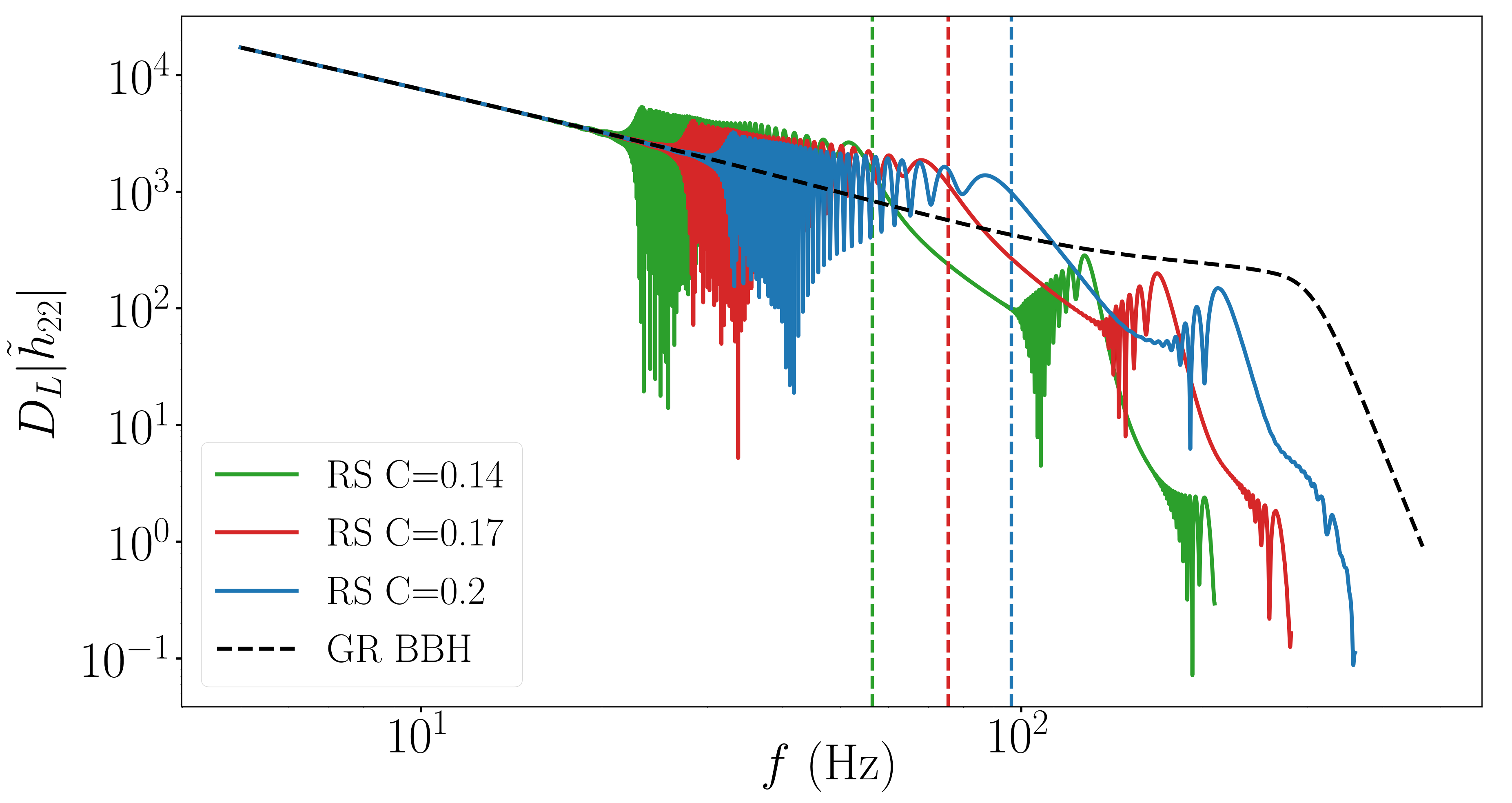}
    }
   \caption{Amplitude of the frequency-domain signal of the three types of ECO binaries, for different values of the initial compactness. The vertical dashed lines indicate the corresponding value of $2f_c$ for each $C_0$.}\label{amp_fd}   
 \end{figure}

We obtain the frequency domain signal $\tilde{h}_{22}$ by performing a discrete Fourier transform. 
In Fig.~\ref{amp_fd} we display the amplitude of $\tilde{h}_{22}$ for the different types of ECO binaries of component masses $30 \ {M_{\odot}}$. Different colors correspond to different $C_0$, and the dashed lines indicate the corresponding GW frequency at contact. For comparison, we show the amplitude for a GR BBH with the same component masses and zero spins, obtained with the IMRPhenomD inspiral-merger-ringdown model \cite{Husa:2015iqa,Khan:2015jqa}, in black dashed lines. Each type of ECO binary exhibits characteristic features in its signal: 

\begin{itemize}
 \item For \emph{NRS} systems (upper panel), $|\tilde{h}_{22}|$ presents one main peak at the frequency of radial 
oscillations, and smaller peaks at the corresponding higher harmonics.

\item For \emph{RBH} systems (middle panel), we observe a main peak at the same frequency around which $\hat{f}$ oscillates in Fig.~\ref{wvf_td_bh}. Beatings between orbital and radial frequencies lead to significant power being emitted at higher and lower frequencies too. As a consequence, the inspiral and post-contact signal ``interfere'', leading to wiggles around the transition frequency $2f_c$. The beatings (and therefore the ``interference'') and the amplitude of the peak are reduced for more compact configurations, because the post-contact signal is shorter. Moreover, as the compactness increases the signal becomes more and more similar to a BBH signal.  

\item For \emph{RS} systems (lower panel), $\hat{f}$ tends to zero (see Fig.~\ref{wvf_td_ns}), thus spreading the peak observed for \emph{RBH} systems over lower frequencies. This leads to strong interferences between the inspiral and the post-contact signal, and a highly oscillatory behavior ensues. We identify the repetition of an oscillatory pattern at different frequencies, corresponding to ``harmonics'' of the signal. 
\end{itemize}


\section{Data analysis}\label{da}

We can now assess the detectability of ECO binaries and how well we could distinguish them from BBHs with different GW detectors. We consider ECO binaries with source frame masses in the ranges $[5 M_{\odot},10^4M_{\odot}]$ and $[10^3 M_{\odot},10^9M_{\odot}]$, which could be probed by ground-based detectors and by the Laser Interferometer Space Antenna (LISA) \cite{Audley:2017drz}, respectively. In the following, it will be convenient to use both \emph{source-frame} (subscript $s$) and \emph{detector-frame} (subscript $d$) masses. The two are related by $m_{0,d}=(1+z)m_{0,s}$, where $z$ is the cosmological redshift. We adopt the cosmology inferred by the Planck mission \cite{Aghanim:2018eyx}. We start by reviewing a few notions of data analysis and describing the specifics of the detectors.

\subsection{Definitions}

The strain measured by ground-based detectors can be written as
\begin{align}
 h=\mathcal{F}_+(\theta,\varphi,\psi,\iota)h_+ + \mathcal{F}_{\times}(\theta,\varphi,\psi,\iota)h_{\times}, \label{strain}
\end{align}
where $\mathcal{F}_+$ and $\mathcal{F}_{\times}$ are the extended antenna pattern functions of the detector, 
including the dependence on the inclination angle ($\iota$) in addition to the declination ($\theta$), right ascension ($\varphi$) and polarization ($\psi$) angles. 
For a given GW detector, we define the inner product between two templates $h_1$ 
and $h_2$ as \cite{Cutler_1994}
\begin{align}
 (h_1|h_2)=2 \int_{f_{\rm min}}^{f_{\rm max}} 
\frac{\tilde{h}_1(f)\tilde{h}^*_2(f)+\tilde{h}^*_1(f)\tilde{h}_2(f)}{S_n(f)} \ {\rm d}f, 
\label{inner_product}
\end{align}
where $S_n(f)$ is the power spectral density (PSD) of the detector, and the integration limits $f_{\rm min}$ and $f_{\rm max}$ depend on the detector and the signal. The overlap is defined as
 \begin{equation}
  \mathcal{O}(h_1,h_2)= \frac{(h_1|h_2)}{\sqrt{(h_1|h_1)(h_2|h_2)}}. \label{olap}
 \end{equation}
and the SNR as
\begin{equation}
 {\rm SNR}=\sqrt{(h|h)}. 
\end{equation}
We define the averaged SNR over $\theta$, $\phi$, $\psi$ and $\iota$ as
\begin{align}
 <{\rm SNR}^2>= \int_{f_{\rm min}}^{f_{\rm max}} 
\frac{(<\mathcal{F}^2_+>+<\mathcal{F}^2_{\times}>)|\tilde{h}_{+,\times}(f)|^2}{
S_n(f) } \ {\rm d}f , \label{snr_av}
\end{align}
where $<>$ denotes averaging over the angles.
Defining $n$ as the number of independent detectors and $\alpha$ as the angle between the arms of a given detector, we have
\begin{align}
 <\mathcal{F}^2_+>&=\frac{7}{75}n \sin^2(\alpha), \label{av_fp}   \\
  <\mathcal{F}^2_{\times}>&=\frac{1}{15}n \sin^2(\alpha). \label{av_fc}
\end{align}
There is no contribution from polarization mixing 
because \mbox{$<\mathcal{F}_+\mathcal{F}_{\times}>=0$}.



\subsection{Detectors}

We consider three ground-based detectors: LIGO Livingston at the time of \emph{GW170817} \cite{Abbott_2017}, advanced LIGO (aLIGO) at its design sensitivity \cite{TheLIGOScientific:2014jea} and the Einstein telescope (ET) \cite{Hild_2011,Punturo:2010zz,Ballmer:2015mvn}. We take respectively $f_{\rm min}=23, 10, 5 \  {\rm Hz}$ and $f_{\rm max}= 2000,3000,8000 \  {\rm Hz}$ for the integration limits. LIGO Livingston and aLIGO are single detectors with angle between the arms $\alpha=\pi/2$, while ET is made of three detectors with $\alpha=\pi/3$.

For LISA, the expression given in Eq.~\eqref{strain} is valid only in the long-wavelength approximation \cite{Cutler:1997ta}, to which we will stick. LISA can then be seen as consisting of two independent detectors with $\alpha=\pi/3$. For more details on the calculation of angle-averaged SNRs in LISA, see~\cite{Cornish:2018dyw}. We assume a 
mission duration of $4 \ {\rm years}$ and adopt the frequency range 
$f_{\rm min}={\rm max}(10^{-5} \  {\rm Hz}, f_{\rm -4 yrs})$ and $f_{\rm max}= 0.1 \  {\rm Hz}$, where $f_{\rm -4 yrs}$ is the frequency of a BBH of same component masses 4 years before merger. This is an optimistic choice,
as it allows for observing the inspiral during the whole mission duration, as well as the merger if it happens in band.

\subsection{Detectability and distinguishability}

  \begin{figure}[th!]{
    \centering \includegraphics[scale=0.09]{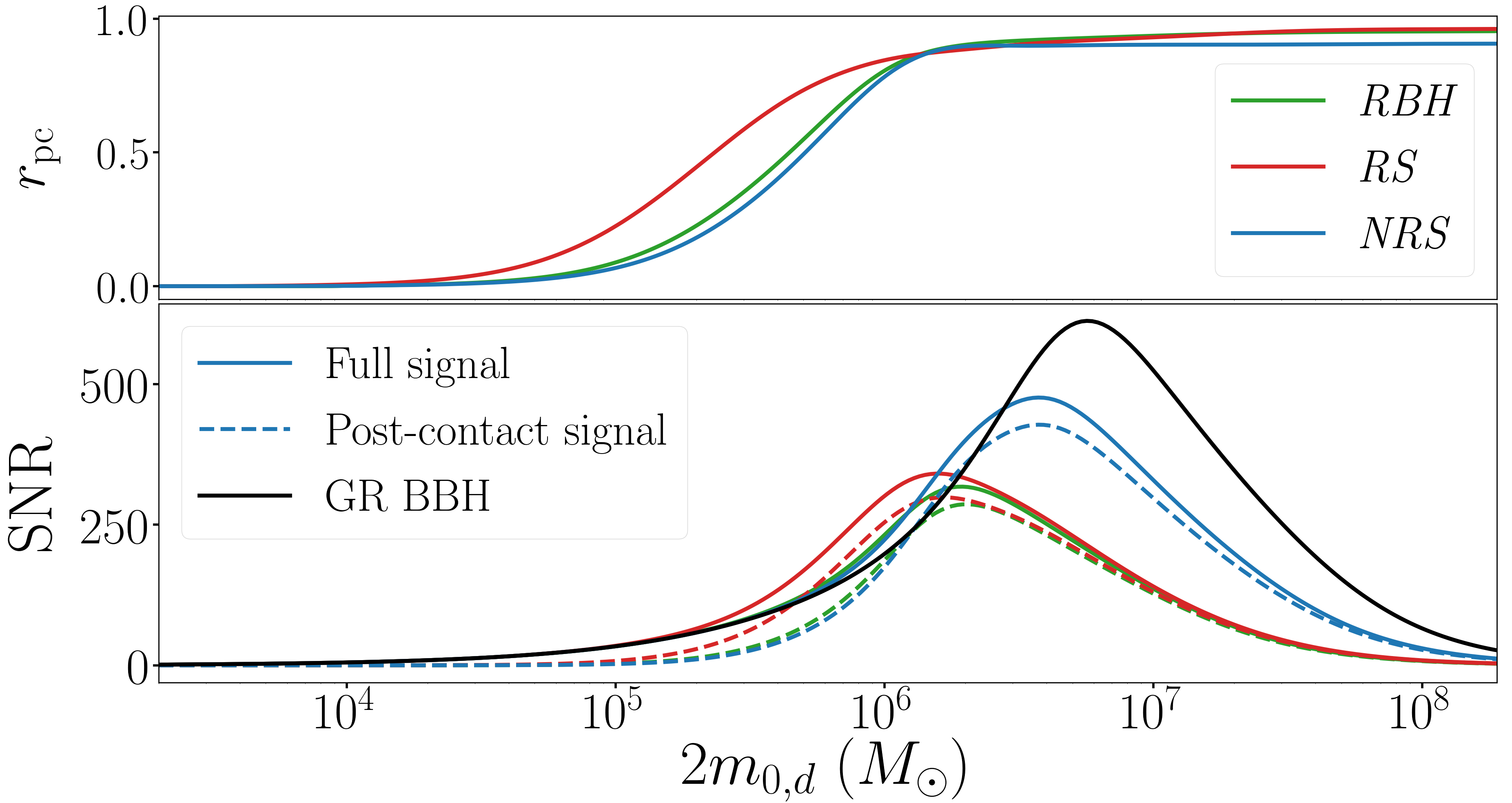}
    }
    \caption{Lower panel: total and post-contact SNR in LISA as a function of the detector-frame mass, for the three types of ECO binaries. Lower panel: ratio of post-contact to total SNR. We take $C_0=0.17$ and place the sources at $z=10$. As the total mass increases, LISA becomes more sensitive to the post-contact stage and less sensitive to the inspiral, and thence $r_{\rm pc}$ approaches unity.}\label{frac_snr}
 \end{figure}

  \begin{figure}[th!]
\subfigure[aLIGO could detect ECO binaries up to $z\simeq 1$. ]{
    \centering \includegraphics[scale=0.09]{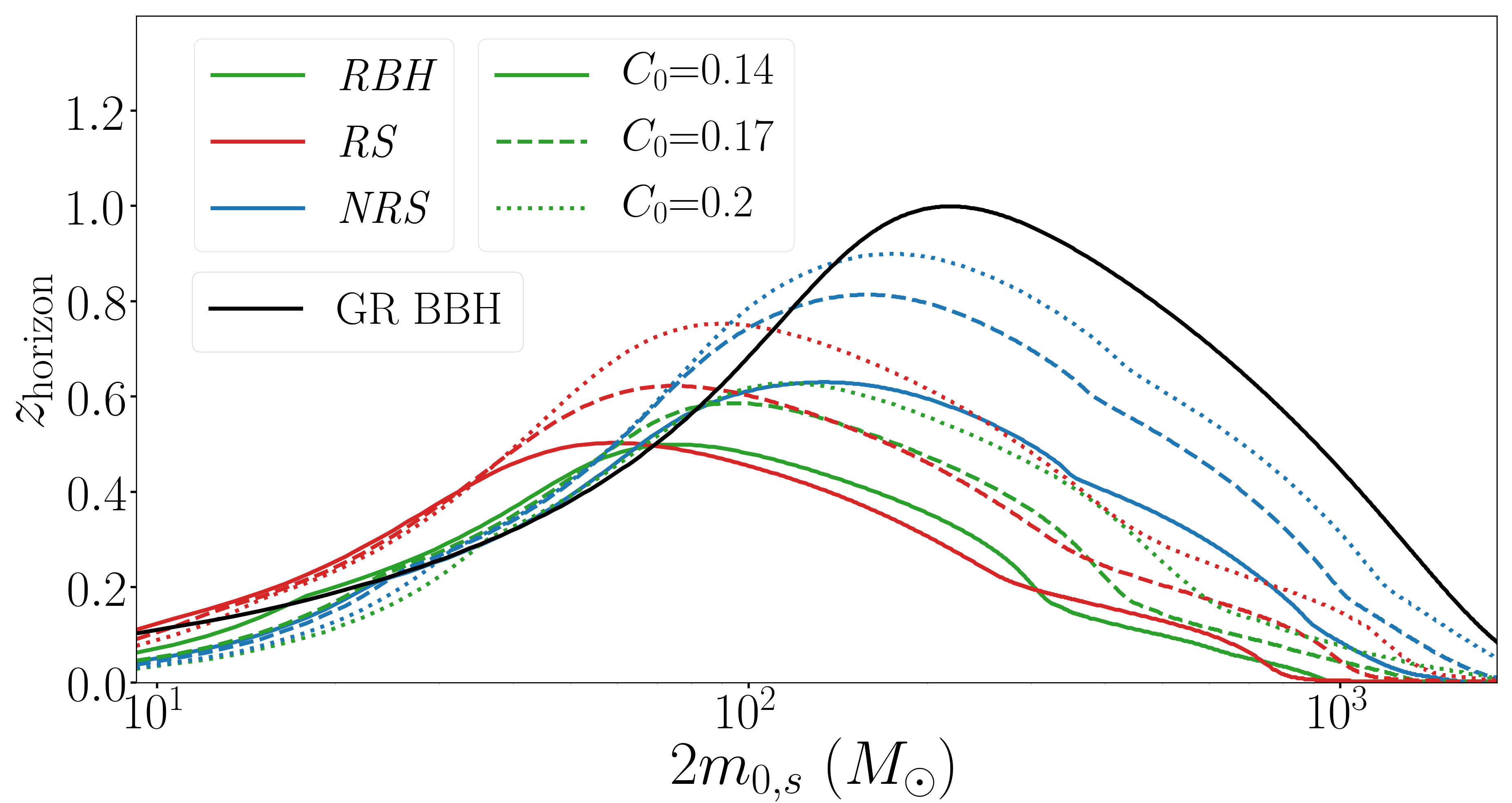}
    }
\subfigure[ET could detect ECO binaries with total mass $\mathcal{O}(10^2) \ M_{\odot}$ throughout the observable Universe. The double-peaked feature for the \emph{NRS} system is due to the presence of two lines in the PSD, which damp the contribution of the secondary peak of the frequency-domain signal for masses in between the two maxima.]{
    \centering \includegraphics[scale=0.09]{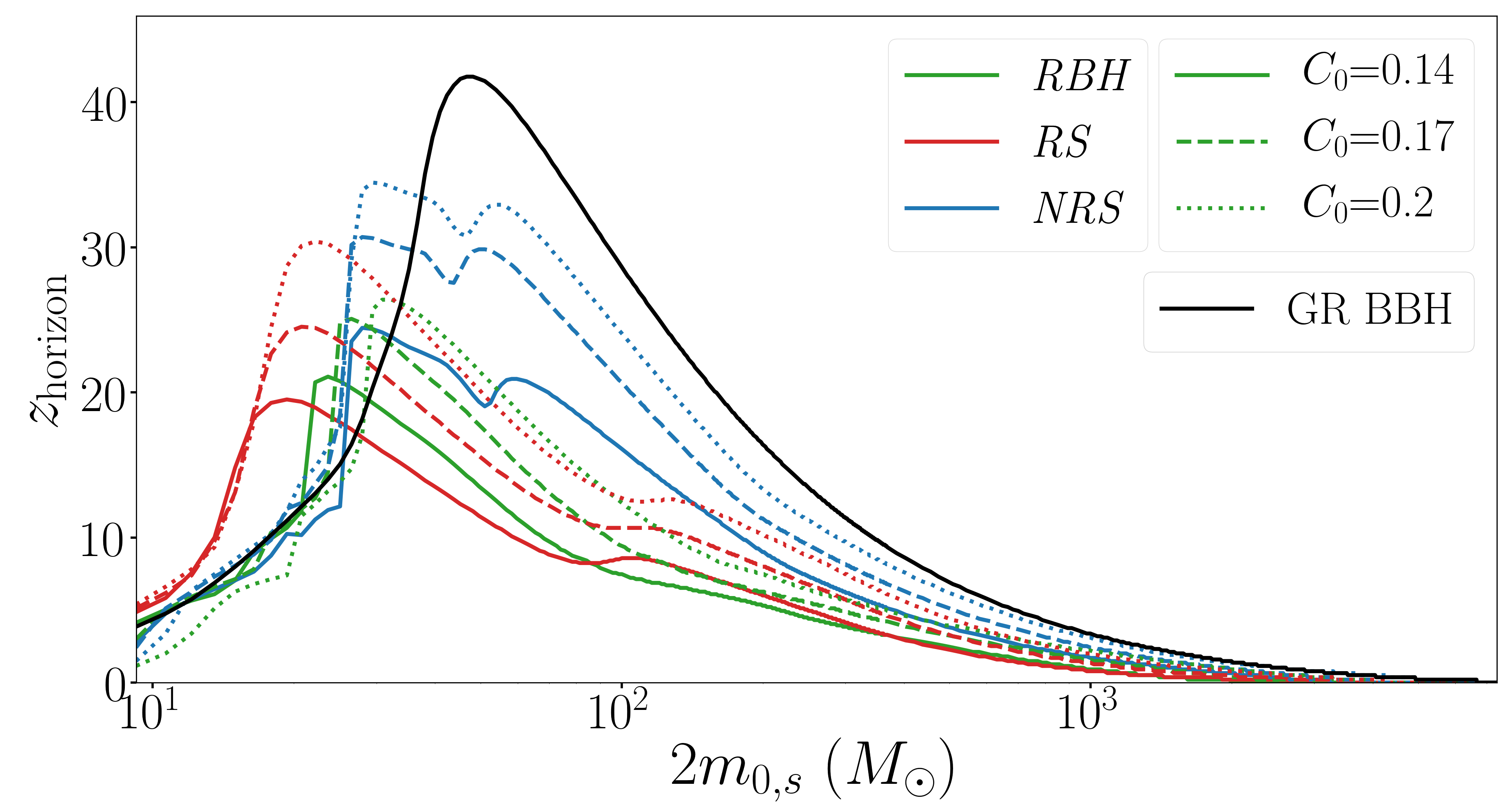}
   }
   \subfigure[LISA could detect ECO binaries with total mass $10^4-10^6 \ M_{\odot}$ throughout the observable Universe. The abrupt cut is due to systems that merge outside the LISA frequency band, but which accrue a large SNR during their inspiral.]{
    \centering \includegraphics[scale=0.09]{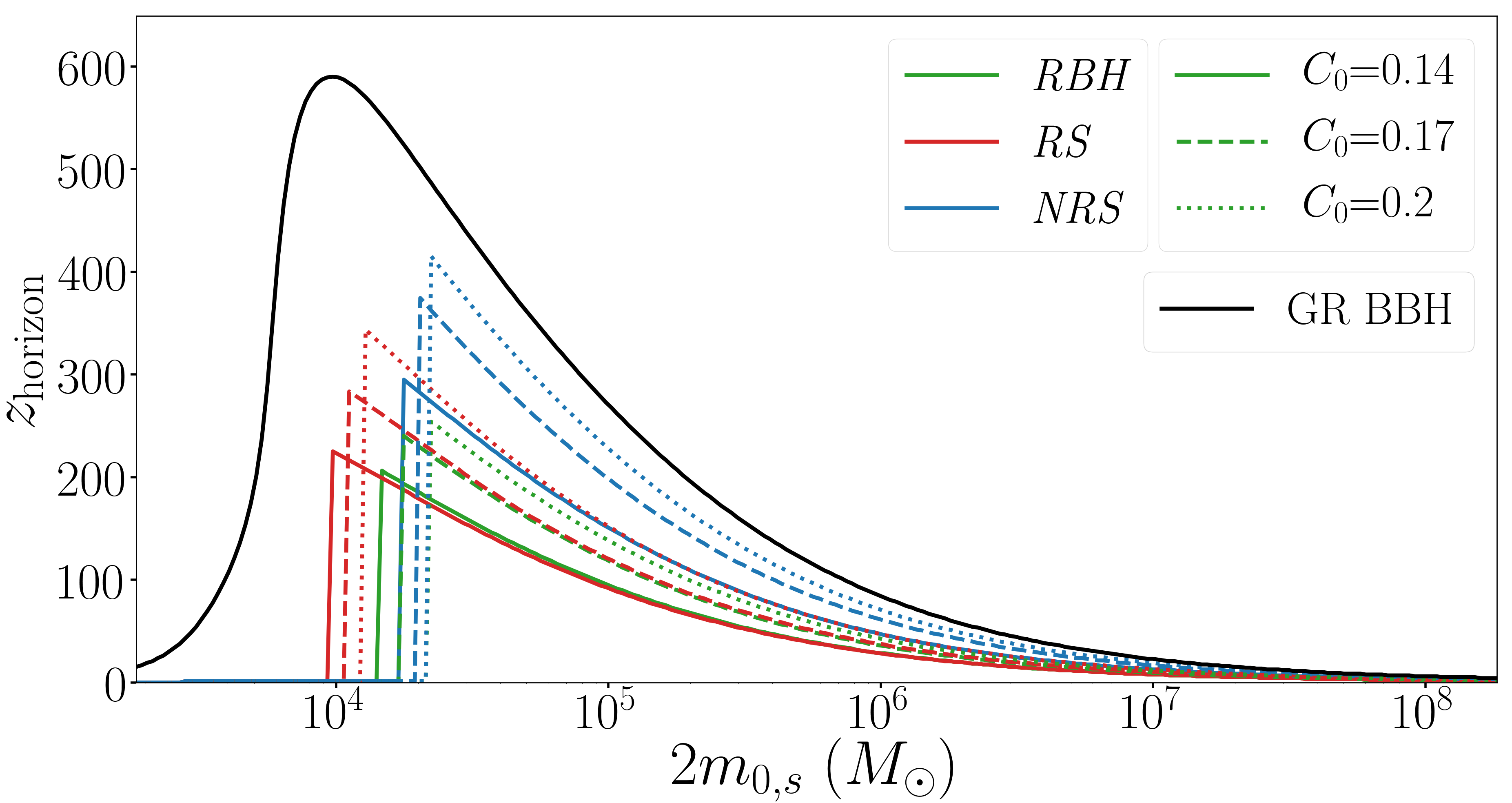}
   }
   \caption{Maximum redshift up to which we could observe and potentially distinguish different types of ECO binaries from BBHs, as a function of the total mass in the source frame. The threshold for observation and distinguishability is set at a total SNR of 8 and at a post-contact SNR of 4, respectively. The maximum redshift typically increases with $C_0$ and, among ECO binaries, is largest for \emph{NRS} systems.
}\label{fig_snrs}   
 \end{figure}

 We start by assessing the fraction of SNR coming from the post-contact signal. We define it as
\begin{align}
 r_{\rm pc}&=\sqrt{\frac{<(\tilde{h}_{\rm pc}|\tilde{h}_{\rm pc})>}{<(\tilde{h}_{\rm tot}|\tilde{h}_{\rm tot})>}} =\frac{SNR_{\rm pc}}{SNR_{\rm tot}},
\end{align}
where $\tilde{h}_{\rm tot}$ is the Fourier transform of the whole signal, and $\tilde{h}_{\rm pc}$ is that of the post-contact signal only. 
In the lower panel of Fig. \ref{frac_snr}, we display the total (full line) and post-contact (dashed line) SNR in LISA, for the three types of ECO binaries and as a function of the total mass in the detector frame. For comparison, in black we plot the SNR for a binary of nonspinning GR BHs, computed with IMRPhenomD. We place the sources at $z=10$. In the upper panel we plot $r_{\rm pc}$, which is independent of the redshift. As the total mass increases, LISA becomes more sensitive to the post-contact stage and less to the inspiral, thus $r_{\rm pc}$ approaches unity. The difference between the various types of systems can be understood from Fig.~\ref{amp_fd}. The post-contact evolution of \emph{RS} systems starts early (at lower frequencies), and the maximum of the SNR is shifted to lower masses than for the other systems. \emph{NRS} systems can reach higher SNRs, because the amplitude of their frequency-domain signal is almost as high as for GR BBHs up to $\simeq 2f_c$, thus more SNR is accumulated in this frequency range as compared to \emph{RS} and \emph{RBH} systems, and significant power is emitted at the harmonics of the radial oscillation frequency (more than by BBHs at the same frequency). For lower masses, the total SNR of ECO binaries can even be larger than for BBHs, but it is smaller for higher masses because of the lower emitted power at high frequencies.
The picture is similar for ECO binaries in the range $[5 M_{\odot},10^4M_{\odot}]$ in ground-based detectors.

To gauge our ability to identify these signals, i.e. to detect them and potentially distinguish them from BBHs, we define two thresholds: one for the whole signal and another one for the post-contact signal.
Based on studies of the detectability of post-merger signals from BNS coalescence \cite{Tsang:2019esi,Chatziioannou:2017ixj,Breschi:2019srl}, we require a minimum SNR for the post-contact signal of 4. Whereas for the overall detectability, we assume a threshold of 8. We define the horizon redshift for the identification of ECO binaries as the maximum redshift such that both thresholds are exceeded. We expect that if both thresholds are exceded, we should be able to spot the presence of an ECO binary's post-contact signal in the residuals left after subtracting the best fit GR BBH template from data, and since it is very different from the post-merger signal of BBHs, we should be able to identify the merging objects as being ECOs. Fig.~\ref{fig_snrs} shows the horizon redshift for ECO binaries, as a function of the total mass in the source frame. The upper panel shows results for aLIGO, the middle one for ET and the lower for LISA. For comparison, we plot in black the horizon redshift for a binary of nonspinning GR BHs, computed with IMRPhenomD. The abrupt cut for LISA is due to systems that accumulate a large SNR during their inspiral, but which merge outside the LISA frequency band. Because they can reach higher SNRs, \emph{NRS} systems have the largest horizon among ECO binaries. The horizon distance typically increases with $C_0$, because the inspiral of more compact binaries lasts longer, allowing for the accumulation of more SNR. Our results show that aLIGO could identify ECO binaries only up to $z\simeq 1$, whereas ET and LISA could identify binaries with total mass $\mathcal{O}(10^2) \ M_{\odot}$ and $10^4-10^6 \ M_{\odot}$ respectively, throughout the observable Universe.

 \begin{figure}[th!]
\subfigure[FFs for \emph{RBH} systems. Because aLIGO observes only the very end of the inspiral, where tidal effects are more pronounced, and the post contact-stage is not ``sufficiently different'' from BBHs, the minimum FF is lower in aLIGO than in ET.]{
    \centering \includegraphics[scale=0.09]{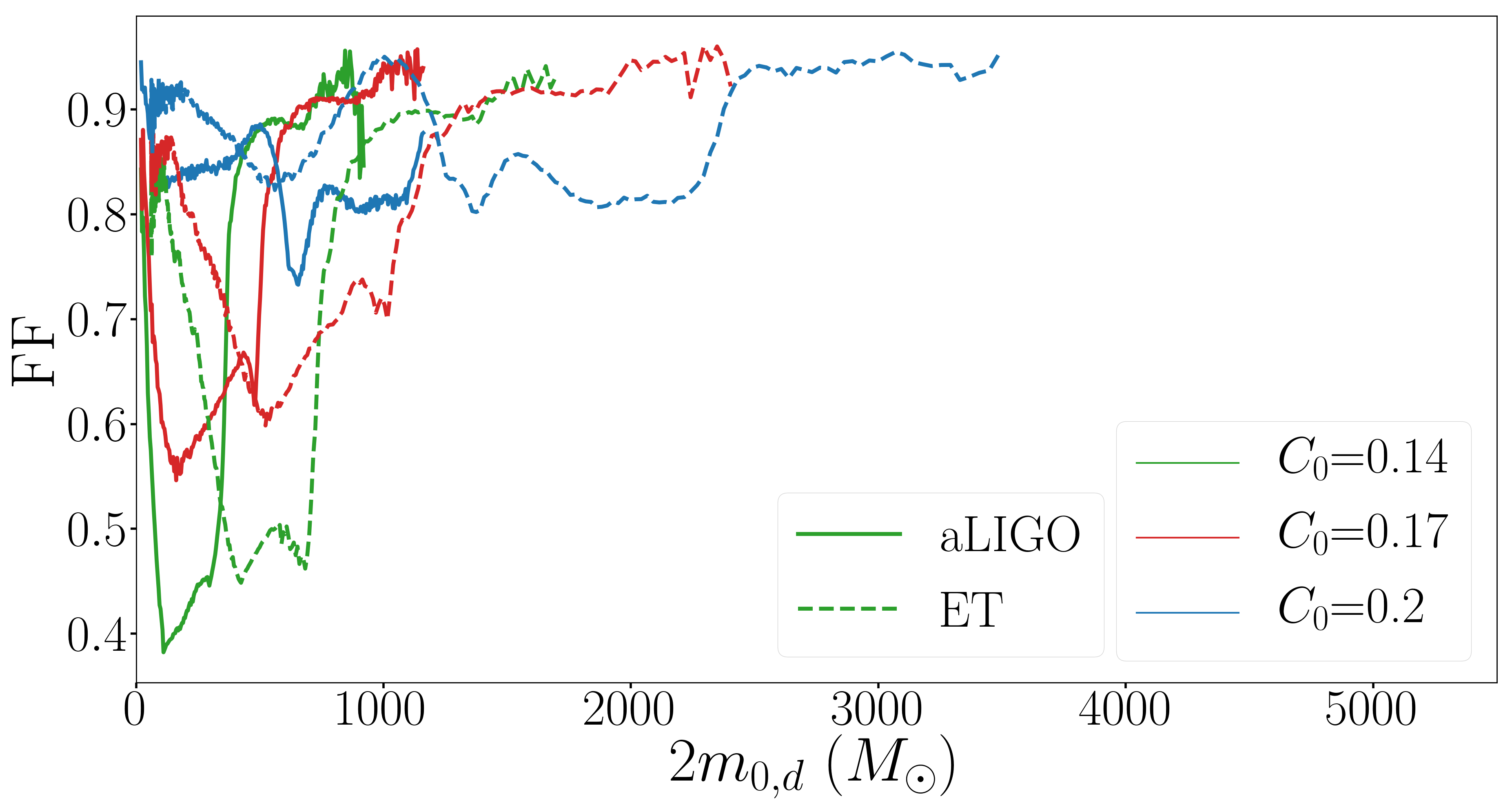}
    }
\hspace{2em}
\subfigure[FFs for \emph{NRS} systems. The post-contact stage is ``sufficiently different'' from BBHs to compensate for the better match in the inspiral with ET. Thus, the minimum FF is lower in ET than in aLIGO, unlike the \emph{RBH} systems shown in the upper panel.]{
    \centering \includegraphics[scale=0.09]{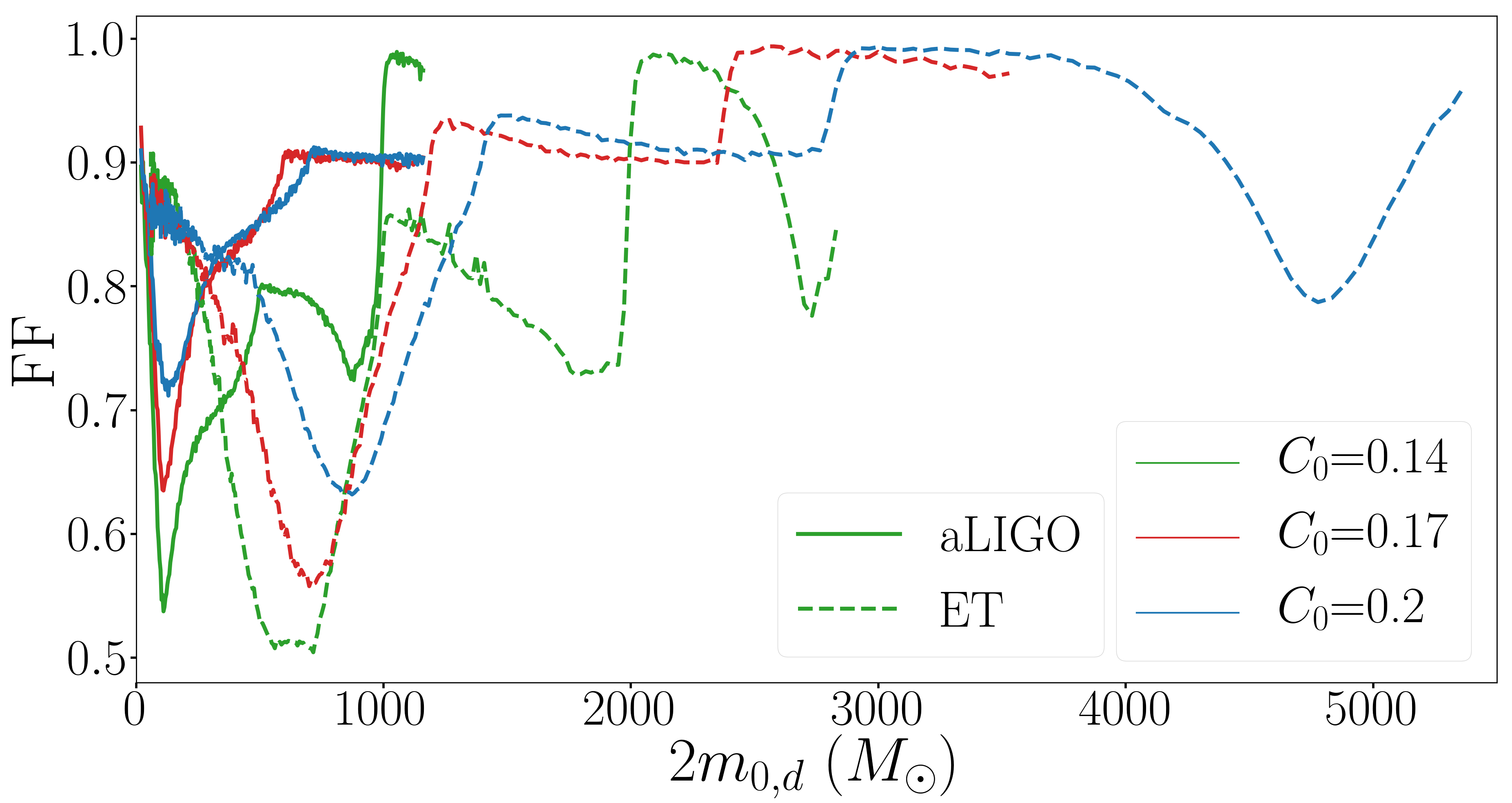}
   }
   \subfigure[FFs for \emph{RS} systems. Since the post-contact stage is ``even more different'' from BBHs than for \emph{NRS} systems, the minimum FF is smaller in both detectors and for all values of $C_0$ than for \emph{NRS} and \emph{RBH} systems.]{
    \centering \includegraphics[scale=0.09]{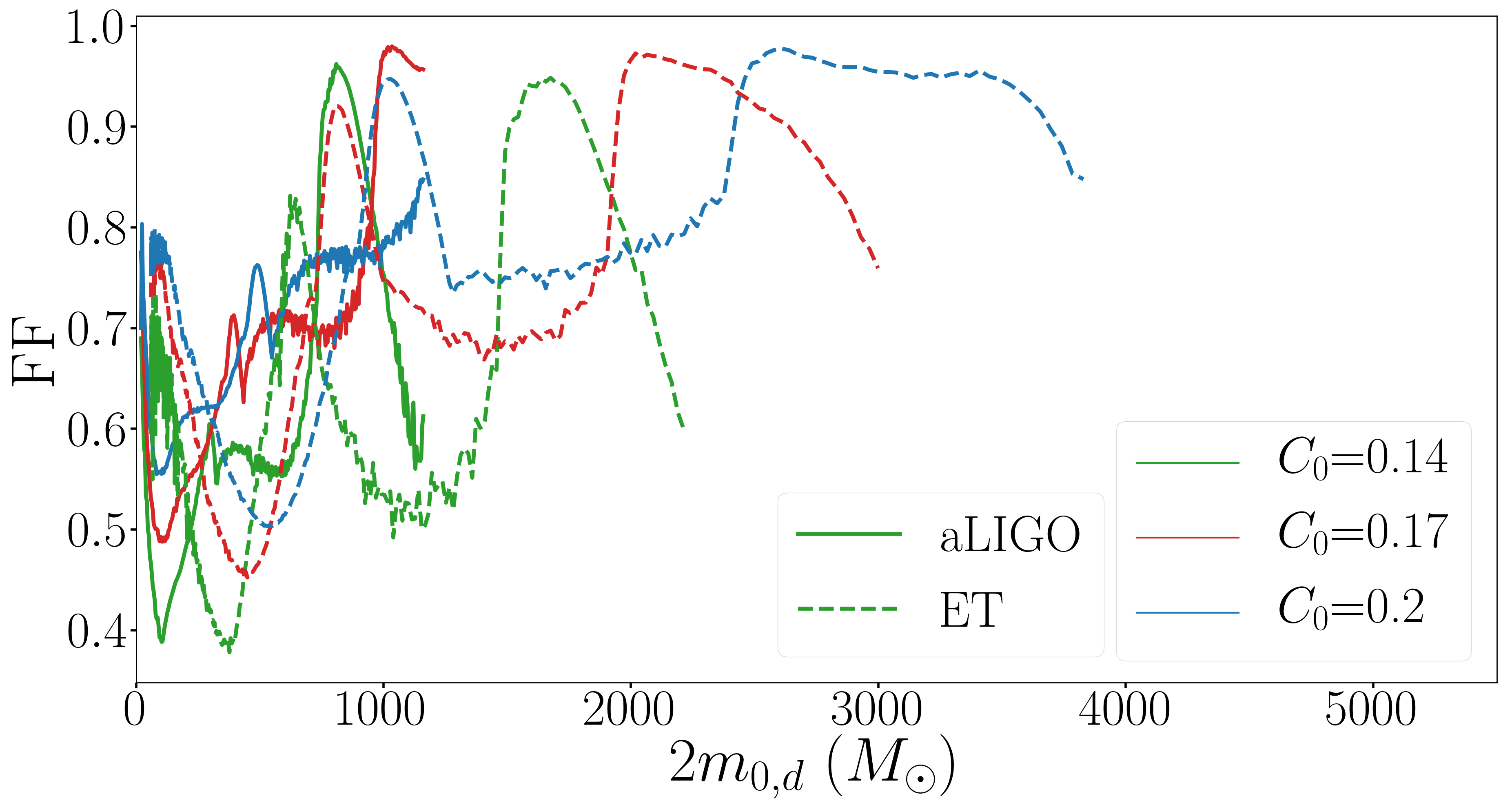}
    }
   \caption{FF as a function of the total mass in the detector frame, for the different types of ECO binaries (one on each panel) and different values of $C_0$ (distinguished by the color) in aLIGO (full lines) and ET (dashed lines). Up to $60\%$ of the SNR could be lost if ECO binaries are detected with BBH templates, potentially jeopardizing detection of weaker signals.}\label{ffs}   
 \end{figure}

\begin{table*}[t]
\centering
\begin{tabular}{|*{13}{c|}}
	\hline
	\multicolumn{1}{|c|}{\multirow{2}{*}{Event}} &  \multicolumn{1}{c|}{\multirow{2}{*}{$\mathcal{M}_{c,s} \ (M_{\odot})$}} &  \multicolumn{1}{c|}{\multirow{2}{*}{$m_{0,s} \ (M_{\odot})$}} & \multicolumn{1}{c|}{\multirow{2}{*}{${\rm SNR_{obs}}$}}  & \multicolumn{3}{c|}{$C_0=0.14$} & \multicolumn{3}{c|}{$C_0=0.17$} & \multicolumn{3}{c|}{$C_0=0.2$}\\ \cline{5-13}
	& & & & \emph{RBH} & \emph{RS} & \emph{NRS} & \emph{RBH} & \emph{RS} & \emph{NRS} & \emph{RBH} & \emph{RS} & \emph{NRS}  \\ \hline
	\emph{GW150914} & $28.6$ & $33$  & $24.4$ & $22.1$  & $21.7$  & $21.9$  & $20.5$ & $21.1$ & $20.2$  & $17.1$ & $20.1$ & $18.2$ \\ 
	\emph{GW150112} & $15.2$ & $17$  & $10.0$ & $8.0$   & $8.3$   & $7.2$   & $6.9$  & $8.0$  & $6.5$   & $5.2$  & $7.5$  & $5.6$  \\ 
	\emph{GW151226} & $8.9$  &  $10$ & $13.1$ &  $7.4$  & $9.4$   & $6.0$   & $5.9$  &  $8.6$ & $5.2$   & $4.0$  & $7.8$  & $4.3$ \\ 
	\emph{GW170104} & $21.4$ & $25$  & $13.0$ &  $11.2$ & $11.3$  & $11.0$  & $10.3$ & $11.0$ & $10.1$  & $8.3$  & $10.4$ & $8.9$ \\ 
	\emph{GW170608} & $7.9$  &  $9$  & $14.9$ & $7.6$   &  $10.1$ & $6.0$   & $5.9$  & $9.2$  & $5.1$   & $4.0$  & $8.3$  & $4.2$ \\ 
	\emph{GW170729} & $35.4$ & $41$  & $10.8$ & $10.3$  & $10.4$  & $10.4$  & $10.0$ & $9.8$  & $9.9$   & $8.8$  & $9.3$  & $9.1$  \\ 
	\emph{GW170809} & $24.9$ & $29$  & $12.4$ & $11.1$  & $11.0$  & $11.0$  & $10.3$ & $10.6$ & $10.2$  & $8.5$  & $10.2$ & $9.1$ \\ 
	\emph{GW170814} & $24.1$ & $28$  & $15.9$ & $13.7$  & $14.0$  & $13.7$  & $12.8$ & $13.4$ & $12.6$  & $10.4$ & $12.8$ & $11.1$  \\ 
	\emph{GW170818} & $26.5$ & $30$  & $11.3$ & $10.3$  & $10.1$  & $10.2$  & $9.6$  & $9.8$  & $9.4$   & $8.0$  & $9.3$  & $8.5$  \\ 
	\emph{GW170823} & $29.2$ & $33$  & $11.5$ & $10.8$  & $10.4$  & $10.8$  & $10.2$ & $10.2$ & $10.1$  & $8.7$  & $9.7$  & $9.2$ \\ \hline

\end{tabular}
 \caption{SNR of the post-contact signal for the LVC events, under the hypothesis that those were
produced by ECO binaries.
We assume equal masses
and use the chirp mass and redshift reported in \cite{LIGOScientific:2018mvr} to compute the masses of the binaries components; we also fix the total SNR to the network SNR reported by the LVC (${\rm SNR_{obs}}$).
}\label{snr_cat}
\end{table*}

Next, we investigate whether the detection of an ECO binary may be mistaken for a BBH. For a given ECO binary signal, we define its fitting factor (FF) \cite{Apostolatos:1995pj} as the maximum overlap over GR BBH templates. The FF measures the \emph{effectualness} \cite{Damour:1997ub} of a family of templates at reproducing a fiducial signal. The difference 1-FF yields the SNR fraction that would be lost as a result of the mismatch between the signal and the (best) template. For comparison, template banks are built so that the overlap between neighboring templates is no less than 0.97 \cite{Colaboration:2011np,Aasi:2012rja}. For simplicity, we neglect the dependence on the inclination, declination, right ascension and polarization angles and work with the dominant 22-mode. Moreover, we restrict ourselves to aligned (or anti-aligned) spins, and maximize over the merger time and the phase at merger as in \cite{Allen:2005fk}. Thus, we are left with four parameters over which to maximize: the masses and the spins. For this last step, we use \texttt{Multinest} \cite{Feroz:2008xx} to search for the highest FFs. We compute BBH templates with the IMRPhenomD model.

Fig.~\ref{ffs} shows FFs computed for each type of ECO binary. We use full (dashed) lines for aLIGO (ET). The key information on this figure is given by the range of FFs and the masses for which the FF is minimum. We find that up to $60\%$ of the SNR could be lost if only BBH templates were used in template-based searches, drastically decreasing our chances of detecting these exotic signals. The FF closely resembles inverted  SNR (or equivalently the horizon redshift) as a function of the total mass (see Fig.~\ref{fig_snrs}), and is minimum for masses that maximize the post-contact SNR. This is why the minimum FF for ET is displaced with respect to aLIGO. The oscillations at higher masses are due to the most salient parts of these exotic signals lying within the most sensitive frequency window of the detector. \emph{RS} and \emph{NRS} systems have lower FFs in ET, whereas for \emph{RBH} systems the minimum FF is lower in aLIGO. This might seem counterintuitive, but it can be understood as follows: (i) aLIGO observes only the very end of the inspiral, where tidal effects are more pronounced and dephasing with BBH templates grows fast; (ii) the post-contact signal of \emph{RS} and \emph{NRS} systems is ``sufficiently different'' from BBHs, unlike that of \emph{RBH} systems; (iii) the noise level in ET is approximately flat over a broad frequency band, thus ET is more sensitive to the post-contact stage. Similarly, the FF is overall smaller for \emph{RS} systems and higher for \emph{RBH} systems (in agreement with the visual impression from Fig.~\ref{amp_fd}). Finally, the FF increases with $C_0$, because 
ECOs become more similar to BHs as their compactness increases. Since the \emph{effectualness} is always larger than the \emph{faithfulness} \cite{Damour:1997ub}, our results suggest that the estimation of the parameters of these exotic sources (including masses, spins, etc) could be significantly biased if only GR BBH templates are used.

Finally, we investigate the possibility that the detected BBH events in the first GW catalog released by the LVC \cite{LIGOScientific:2018mvr} may actually be ECO binaries. 
Because the chirp mass is one of the best constrained parameters, we make the assumption that its measurement is not significantly biased, and use the values reported by the LVC. Moreover, for each event we assume equal masses (all events in the catalog are compatible with this assumption). We estimate the SNR of the post-contact signal by fixing the total SNR to the value measured by the LIGO/Virgo network, and by using the sensitivity of LIGO Livingston at the time of \emph{GW170817}. Let us stress that the noise level was not the same for all the events, and in particular, the PSD at the time of \emph{GW170817} corresponds to the highest sensitivity reached during the first two observing runs of the LVC. Therefore, our estimates serve as higher bounds on what the real post-contact SNR would have been.
The values are reported in Table \ref{snr_cat}. The post-contact SNR decreases with $C_0$, because the SNR accumulated during the inspiral is larger for more compact binaries and we have fixed the total SNR. Based on studies of the detectability of the post-contact signal of BNSs \cite{Chatziioannou:2017ixj}, our results suggest that the post-contact SNR would be sufficient
for detection with wavelet-based pipelines, at least for the loudest sources. The compatibility of the residuals left after subtracting the best fit GR BBH template from data with Gaussian noise \cite{LIGOScientific:2019fpa} makes it unlikely that these events were generated by ECO binaries as the ones considered in this paper (i.e. made of identical objects with $C_0 \leq 0.2$ and which do not collapse promptly following contact).

\section{Discussion}\label{ccl}

In the next few years, the sensitivity of ground-based GW detectors will improve significantly, increasing the detection rate and our ability to extract the
source parameters. Moreover, LISA will allow one to observe a yet unexplored population of massive compact binaries with very high SNR (see e.g.~\cite{Sesana:2004sp,Sesana:2004gf,Klein:2015hvg,Bonetti:2018tpf,Katz:2019qlu,Dayal2019,Barausse:2020mdt}). Together, ground-based and space-based detectors will offer us the opportunity to probe the nature of compact objects with exquisite precision. To enhance our chances of detecting exotic signals and properly extracting physical information from the data, we have proposed a simple model that captures the main features of the full GW signal from binaries consisting of identical ECOs with compactness below $0.2$ and which do not collapse promptly following contact. We have focused on the BNS and BBS examples to build our model, since these are the only non-BH binary systems for which numerical simulations are available. Nevertheless, the physics that enters the phenomenological description that we propose is sufficiently generic that we expect it to encompass a wide variety of ECO binaries. 

We adopt an agnostic approach and distinguish three types of ECO binaries, according to their post-contact evolution. We model the post-contact dynamics with a toy model inspired by knowledge of the coalescence phase of compact objects other than BHs. We then extract the GW signal emitted during the post-contact evolution and match it to the inspiral. In the case of collapse to a BH, we also attach a ringdown signal. Our model can qualitatively reproduce the main features of non-BBH signals observed in numerical simulations.

We have investigated the possibility of detecting these objects and discriminating them from BHs with current and future GW detectors. We find that ET and LISA will allow one to detect and potentially distinguish exotic binaries from BBHs (with total mass $\mathcal{O}(10^2) \ M_{\odot}$ and $10^4-10^6 \ M_{\odot}$ respectively) throughout the observable Universe, as compared to up to $z \lesssim 1$ for aLIGO. On the other hand, we find that up to $60 \%$ of the SNR could be lost when using BBH templates to search for these exotic signals, thus affecting our chances of observing them. Finally, we have estimated what the post-contact signal would have been like if the events in the first GW catalog released by the LVC were ECO binaries as the ones considered in this paper. We have found that, for the loudest events, the post-contact signal would have been sufficiently strong to be detected by wavelet-based searches, thus making this hypothesis unlikely. Our analysis could be extended to the second catalog released by the LVC \cite{Abbott:2020niy}, including the noteworthy event \emph{GW190521} \cite{Abbott:2020tfl}, which has been suggested to be compatible with a BBS signal \cite{CalderonBustillo:2020srq}.

In this work, we have focused on distinguishing ECO binaries from BBHs. However, our framework should also allow one to distinguish them from BNSs. For instance, for similar masses, BBSs emit GWs at lower frequencies than BNSs during the post-contact stage. By measuring the masses from the inspiral signal, this feature could allow one to distinguish ECO binaries from BNSs. Moreover,  model selection between classes of ECOs may be feasible with our approach. For instance, ECOs retaining a non-negligible amount of orbital angular momentum at merger could be distinguished from those that do not (e.g. BSs). On the other hand, more subtle effects would require further enhancing our model. 

As a follow-up, one could perform a more refined analysis of the exotic signals that we generated, in order to quantify how well 
physical information can be extracted from them. To this purpose, one may use, 
for instance, the methods developed for the analysis of the post-contact signal of BNSs \cite{Tsang:2019esi,Chatziioannou:2017ixj,Breschi:2019srl}.
On the modeling side, in this work we have considered the angular momentum to be conserved, or completely radiated right after contact. We have observed that these prescriptions lead to very distinct features in the GW signal. Therefore, it would be interesting to consider prescriptions between these two extremes. Finally, additional modifications due to the absence of tidal heating \cite{Maselli:2017cmm,Datta:2019epe} and a multipole structure different from BBHs \cite{Krishnendu:2017shb,Krishnendu:2018nqa} may occur in the inspiral of ECO binaries. Accounting for them should increase our chances of discriminating these sources from BBHs \cite{Pacilio:2020jza}. Moreover, one could account for ECOs with negative Love numbers as suggested in \cite{Cardoso:2017cfl}. Moreover, we could extend our framework to objects with higher compactness, by considering different prescriptions for the merger/post-merger stage, e.g. attaching a rotating bar instead of the toy model we used in this work.

We have proposed a model for the coalescence of ECO binaries that could serve to test different data analysis strategies. It could be particularly helpful for the design of algorithms looking for deviations from GR BBHs around the merger. Certainly, the detection of such exotic signals would be an exciting discovery.

\section*{Acknowledgements}
We are grateful to Tallulah Frappier for Fig.~\ref{scheme_eco} and to Kentaro Takami, Luciano Rezzolla and Luca Baiotti for their permission to use Fig.~\ref{fig_tm}. We are indebted to Paolo Pani, Caio Macedo, Frank Ohme, Duncan Brown and Tim Dietrich for fruitful discussions during the preparation of this work and to Abhirup Ghosh and Juan Calderón Bustillo for their constructive comments in the final stages of the elaboration of this manuscript. A.T. is thankful to SISSA and Perimeter Institute for their hospitality. This work has been supported by the European Union's Horizon 2020 research and innovation program under the Marie Sk\l{}odowska-Curie grant agreement No 690904. The authors would also like to acknowledge networking support from the COST Action CA16104. S.B. and A.T. acknowledge support by CNES, in the framework of the LISA mission. E.B. acknowledges financial support provided under the European Union's H2020 ERC Consolidator Grant ``GRavity from Astrophysical to Microscopic Scales'' grant agreement no. GRAMS-815673.
\appendix

\section{Potential}\label{app:tm}

The full gravitational potential, accounting for the interaction between the cores, the one between the cores and the disk and the rest energy of the disk is given by
\begin{equation}
V_{{\rm gravitational}}=-\frac{Gm^2}{8\rho}-\frac{GmM\rho}{R^2}-\frac{2GM^2}{3R} \label{Vgrav}.
\end{equation}.
The centrifugal and spring terms are given by
\begin{align}
 V_{{\rm centrifugal}}&=\frac{1}{2} I \omega^2= \frac{1}{2} \frac{J^2}{\frac{MR^2}{2}+m\rho^2}, \label{Vcent} \\
 V_{{\rm spring}}&=2k (\rho-\rho_0)^2.
\end{align}


\section{Gravitational waves feedback}\label{app:gw}

The conservative equations of motions (for $P_{\rm GW}=\dot{J}_{\rm GW}=0$) are obtained by taking the derivatives of \eqref{eq_rho_gw} and \eqref{eq_omega_gw} and making use of \eqref{eq_e_gw} and \eqref{eq_j_gw}. They read
\begin{align}
  & \ddot{\rho}=\rho \omega^2 - 4 \omega_0^2(\rho -\rho_0)-\frac{Gm^2}{8\rho^2}+\frac{GM}{R^2}, \label{rho_nd} \\
  & \dot{\omega} =-\frac{2m\rho \dot{\rho} \omega}{\frac{MR^2}{2}+ m \rho^2} \label{omega_nd}. 
 \end{align}
The quadrupole momentum of the system is given by
\begin{align}
  M_{ij}= & \frac{m \rho^2}{2} \begin{pmatrix} 
 \cos(2\phi) &  \sin(2\phi) & 0 \\
 \sin(2\phi) & -\cos(2\phi) & 0 \\
0 & 0 & 0
\end{pmatrix} \nonumber \\ + & \frac{m \rho^2}{6}  \begin{pmatrix} 
 1 &  0 & 0 \\
 0 & 1 & 0 \\
0 & 0 & -2
\end{pmatrix}. \label{quad}
 \end{align}
 We call $Q_{ij}$ the transverse traceless part of $M_{ij}$. 
 Making use of the conservative equations of motion, Eqs. \eqref{rho_nd} and \eqref{omega_nd}, we compute the two polarizations 
 \begin{align} 
 h_{+}&=\frac{\ddot{Q}_{11}-\ddot{Q}_{22}}{D_L} \nonumber \\
  &=\frac{2}{D_L}\frac{G}{c^4}(\gamma_2 \cos(2\phi)-\gamma_1 \sin(2\phi)), \\
 h_{\times}&=\frac{2\ddot{Q}_{12}}{D_L} \nonumber \\
 &=\frac{2}{D_L}\frac{G}{c_4}(\gamma_1 \cos(2\phi)+\gamma_2 \sin(2\phi)),
\end{align}
where $\gamma_1$ and $\gamma_2$ have been defined in Eqs. \eqref{def_f1} and \eqref{def_f2}.
The instantaneous power and angular momentum are given by
\begin{align}
 P_{{\rm GW}}&=\frac{G}{5c^5}\dddot{Q}_{ij}\dddot{Q}_{ij}, \\
 \dot{J}_{{\rm GW}}&=\frac{2G}{5c^5}\epsilon^{3kl}\dddot{Q}_{ka}\dddot{Q}_{la}.
\end{align}
Using the conservative equations of motion and defining
\begin{align}
 g_2 =& \dot{\gamma_1}+2\omega \gamma_2 \nonumber \\
 =&-m \rhod  \left ( \frac{4MR^2\omega^2 \rho }{\frac{MR^2}{2}+m\rho^2}+4 \omega_0^2 (4\rho -3\rho_0) + \frac{Gm}{8\rho^2} -\frac{3GM}{R^2}  \right ), \nonumber \\
 g_1 =& \dot{\gamma_2}-2\omega \gamma_1 \\
 =&\frac{2m}{\frac{MR^2}{2}+m\rho^2} \left[- \frac{m\rho^2 \rhod^2 \omega (3MR^2+2m\rho^2)}{\frac{MR^2}{2}+m\rho^2}  \right .  \nonumber \\ 
 & +\omega \left ( \left (\rhod^2-4 \omega_0^2 (\rho-\rho_0)\rho +\frac{GM\rho}{R^2}-\frac{Gm}{8 \rho} \right ) \left (\frac{3MR^2}{2}+2m\rho^2 \right )   \right . \nonumber \\ 
 & \left . \left .  +\rho^2 \omega^2 \frac{MR^2}{2} \right ) \right ],
\end{align}
we get
\begin{align}
 P_{\rm GW}=&\frac{2}{5} \frac{G}{c^5} ( g_1^2+g_2^2), \label{Prad} \\
 {\dot J}_{\rm GW}=&\frac{4}{5} \frac{G}{c^5} \left (\gamma_2 g_1 - \gamma_1 g_2 \right) \label{Jrad}.
\end{align}
The averaged energy and angular momentum loss are recomputed at the beginning of each cycle.

\section{Quasi-eccentric orbit}\label{app:ecc}

Inspired by the equations for eccentric motion, we introduce the phase angle $\chi$ such that
\begin{equation}
 \rho=\frac{p}{1+e\cos(\chi)}. \label{chi} 
\end{equation}
The parameter $p$ and the ``eccentricity' $e$ of the orbit can be computed from the turning points $\rho_+$ and $\rho_-$
\begin{align}
 p&=2 \frac{\rho_+ \rho_-}{\rho_++\rho_-} \label{param} \\
 e&=\frac{\rho_+-\rho_-}{\rho_++\rho_-} \label{ecc}.
\end{align}
Assuming the adiabatic approximation for $e$ and $p$ holds i.e. 
\begin{align}
 \frac{\dot{e}}{e} \ll \frac{1}{T}, \label{adiab_e} \\ 
 \frac{\dot{p}}{p} \ll \frac{1}{T},
\end{align}
the derivative of Eq.~\eqref{chi} gives
\begin{equation}
 \rhod=\frac{pe\sin(\chi)\dot{\chi}}{(1+e\cos(\chi))^2}. 
\end{equation}
The change of sign in $\dot{\rho}$ is accounted for by $\sin(\chi)$ and the evolution of $\chi$ is monotonic. 
The sign of $\dot{\chi}$ is determined by the initial conditions. 
For instance, if $\chi_i >0$, Eq.~\eqref{eq_rho_gw} becomes
\begin{equation}
 \dot{\chi}=\frac{(1+e\cos(\chi))^2}{pe|\sin{\chi}|}\sqrt{\frac{2}{m}} \sqrt{E-V_{eff}(\rho)}.
\end{equation}

The term under the rightmost square root can be written
\begin{equation}
 E-V_{eff}=\frac{-2km(\rho-\rho_+)(\rho-\rho_-)(\rho-\rho_3)(\rho-\rho_4)(\rho-\rho_5)}{\rho(\frac{MR^2}{2}+m\rho^2)}, \label{eq_pot}
\end{equation}
where $\rho_3$, $\rho_4$ and $\rho_5$ are the remaining (possibly complex) roots of $E-V_{{\rm eff}}=0$. Noticing that
\begin{align*}
 \rho-\rho_+=&\frac{-pe(1+\cos(\chi))}{(1-e)(1+e\cos(\chi))}, \\
 \rho-\rho_-=&\frac{pe(1-\cos(\chi))}{(1-e)(1+e\cos(\chi))},
\end{align*}
the equation for $\chi$ can be recast as
\begin{equation}
 \dot{\chi}=\frac{2\omega_0(1+e\cos(\chi))}{\sqrt{1-e^2}}\sqrt{\frac{(\rho-\rho_3)(\rho-\rho_4)(\rho-\rho_5)}{\rho (\frac{MR^2}{2m}+\rho^2)}}. \label{chidot}
\end{equation}
Thus, we evolve numerically Eqs.~\eqref{chi} and \eqref{chidot} rather than Eq.~\eqref{eq_rho_gw}. 

For \emph{RBH} systems, when the energy becomes larger than the effective potential height, the inner turning point ceases to exist so we cannot use Eq.~\eqref{chi} anymore and we turn back to Eq.~\eqref{eq_rho_gw} to describe the final moments of the evolution before the collapse to a BH.

\ 
\section{Quasi-circular orbit}\label{app:circ}

As the orbit ``circularizes'', $e$ becomes very small and the adiabatic approximation [Eq.~\eqref{adiab_e}] ceases to be valid. This happens when the particle reaches the bottom of the potential well: $V_{{\rm eff}}'(\rho)=0$. This condition allows us to express $\omega$ as a function of $\rho$ and we get the new conservative equations of motion
\begin{align}
\rhod=&0,  \\
 \omega^2=&\frac{1}{\rho} \left ( 4 \omega_0^2 (\rho-\rho_0)+\frac{Gm}{8\rho^2}-\frac{GM}{ R^2} \right ). \label{omega_circ}
\end{align}
These expressions are used in Eqs. \eqref{Jrad} and \eqref{Jrad} to write $P_{\rm GW}$ and $\dot{J}_{\rm GW}$ as functions of $\rho$ only.
Accounting for the dissipation of energy and angular momentum, the equation for $\rho$ is given by
\begin{equation}
 \dot{\rho}=\frac{\dot{E}}{\frac{\partial E}{\partial \rho}}=\frac{\dot{J}}{\frac{\partial J}{\partial \rho}}.
\end{equation}
As a sanity check, we verified that the expressions using the angular momentum and the one using the energy give the same equation.
Finally, the equations of motion become
\begin{align}
 \rhod&=  -\frac{32G}{5c^5}  m^2 \omega^2\rho^2 \left ( 4 \omega_0^2 \rho^2 (\rho-\rho_0)+\frac{Gm}{8}-\frac{GM\rho^2}{ R^2} \right )^2 \nonumber \\
 &\times \left [ 2m\omega_0^2\rho^2 \left ( 4(\rho-\rho_0)\rho^2+\rho_0 \left (\rho^2+\frac{MR^2}{2m} \right ) \right ) \right . \nonumber  \\ 
  & \left .  +\frac{GM}{2R^2\rho^2}\left (\frac{MR^2}{2}-3m\rho^2 \right )-\frac{Gm}{16} \left (\frac{3MR^2}{2}-m\rho^2 \right ) \right ] ^{-1}, \\
 \omega&=\sqrt{\frac{1}{\rho} \left ( 4 \omega_0^2 (\rho-\rho_0)+\frac{Gm}{8\rho^2}-\frac{GM}{ R^2} \right )}, \\
 E&=2m\omega_0^2 \left ( \left (\frac{MR^2}{2m}+\rho^2 \right )\left (1-\frac{\rho_0}{\rho} \right )+(\rho-\rho_0)^2 \right ) \nonumber \\
 &+\frac{Gm}{16\rho^3} \left (\frac{MR^2}{2}+m\rho^2 \right )-\frac{GM}{2R^2\rho} \left (\frac{MR^2}{2}+3m\rho^2 \right )-\frac{2}{3}\frac{GM^2}{R}, \\
 J&=(\frac{MR^2}{2}+m\rho^2)\omega. \label{Jcircle}
\end{align}
Note that taking the $M,k \to 0$ limit we recover the equations for a quasicircular binary at separation $2 \rho$. In this regime, $\rho$ tends an equilibrium value corresponding to one of the roots of Eq.~\eqref{omega_circ}, and $\omega$ goes to 0.
To ensure numerical stability, we switch to this description when $e<10^{-5}$.

\section{Post-contact dynamics}\label{app:dyn}

 \begin{figure*}[th!]
\subfigure[The distance to the center of mass oscillates around a decreasing mean value, untill it reaches a value such that $C_{\rho}=0.5$, leading to the formation of a BH.]{
    \centering \includegraphics[scale=0.085]{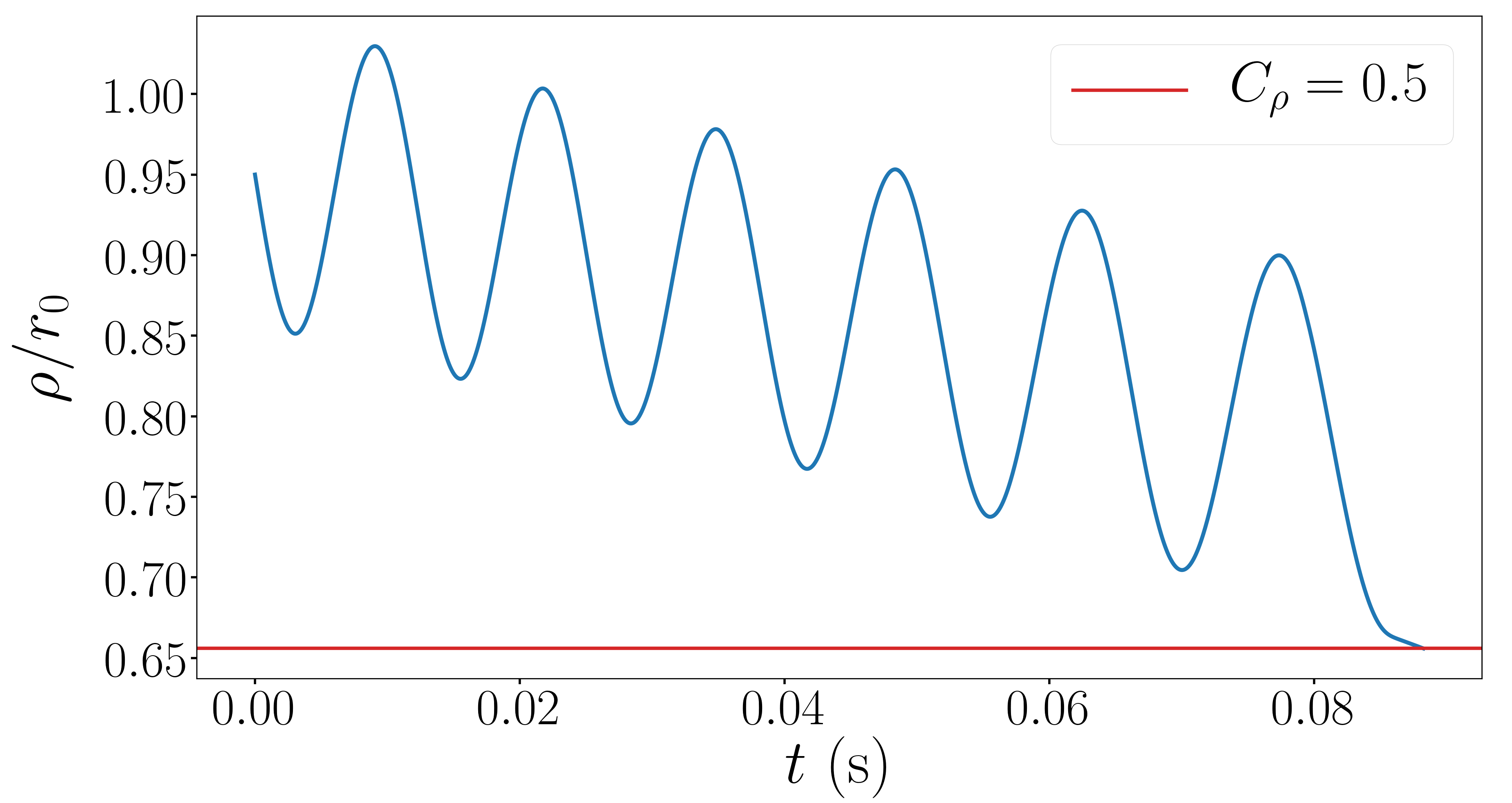}
    }
\hspace{2em}
\subfigure[The orbital angular velocity oscillates around a stable mean value and increases right before the formation of the BH.]{
    \centering \includegraphics[scale=0.085]{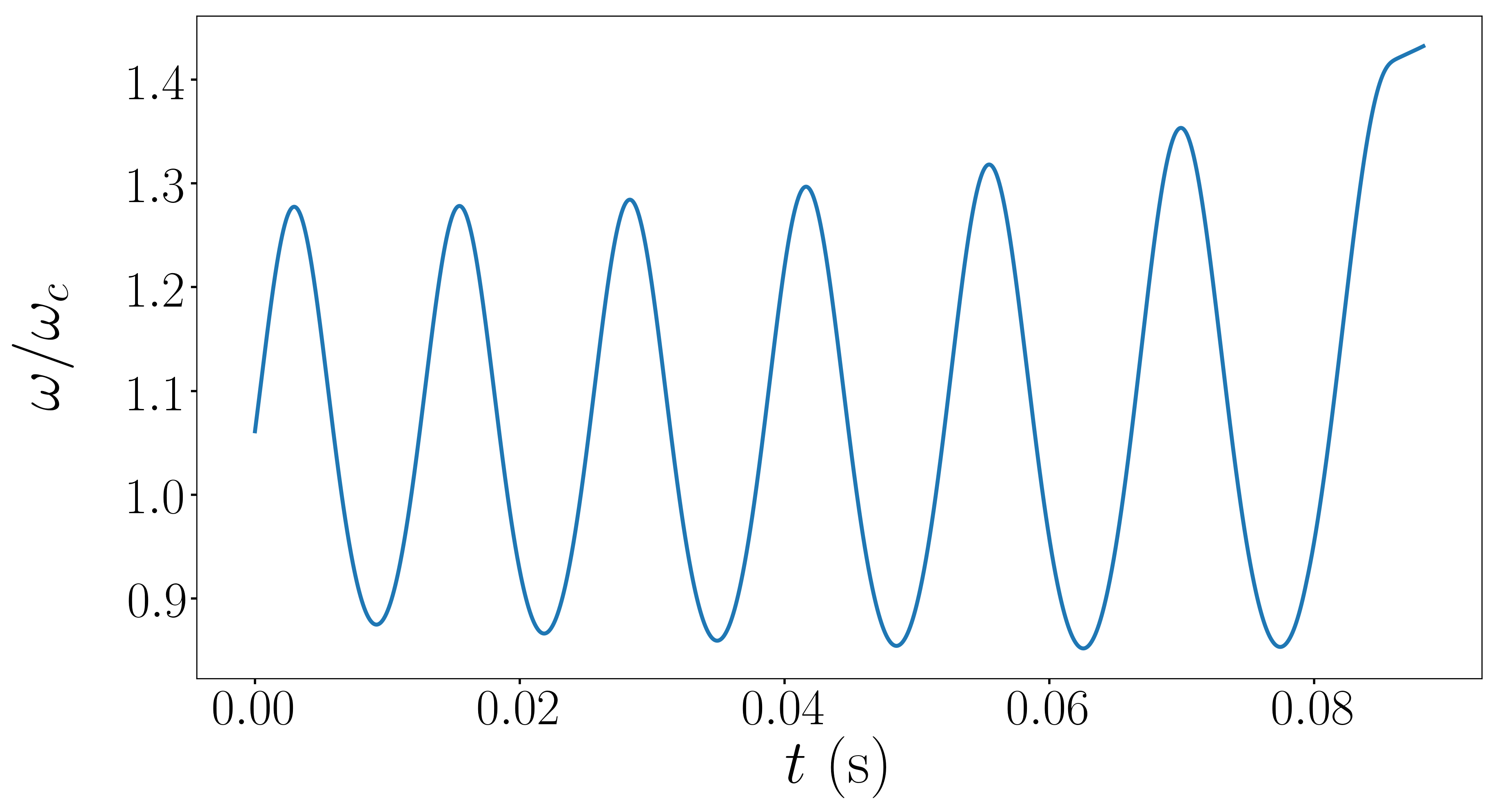}\label{omega_bh}
   }
   \subfigure[Initially, the distance to the center of mass oscillates around a decreasing mean value. Once the orbit ``circularizes'', $\rho$ decreases while tending to an equilibrium value, corresponding to the formation of a stable ECO.]{
    \centering \includegraphics[scale=0.085]{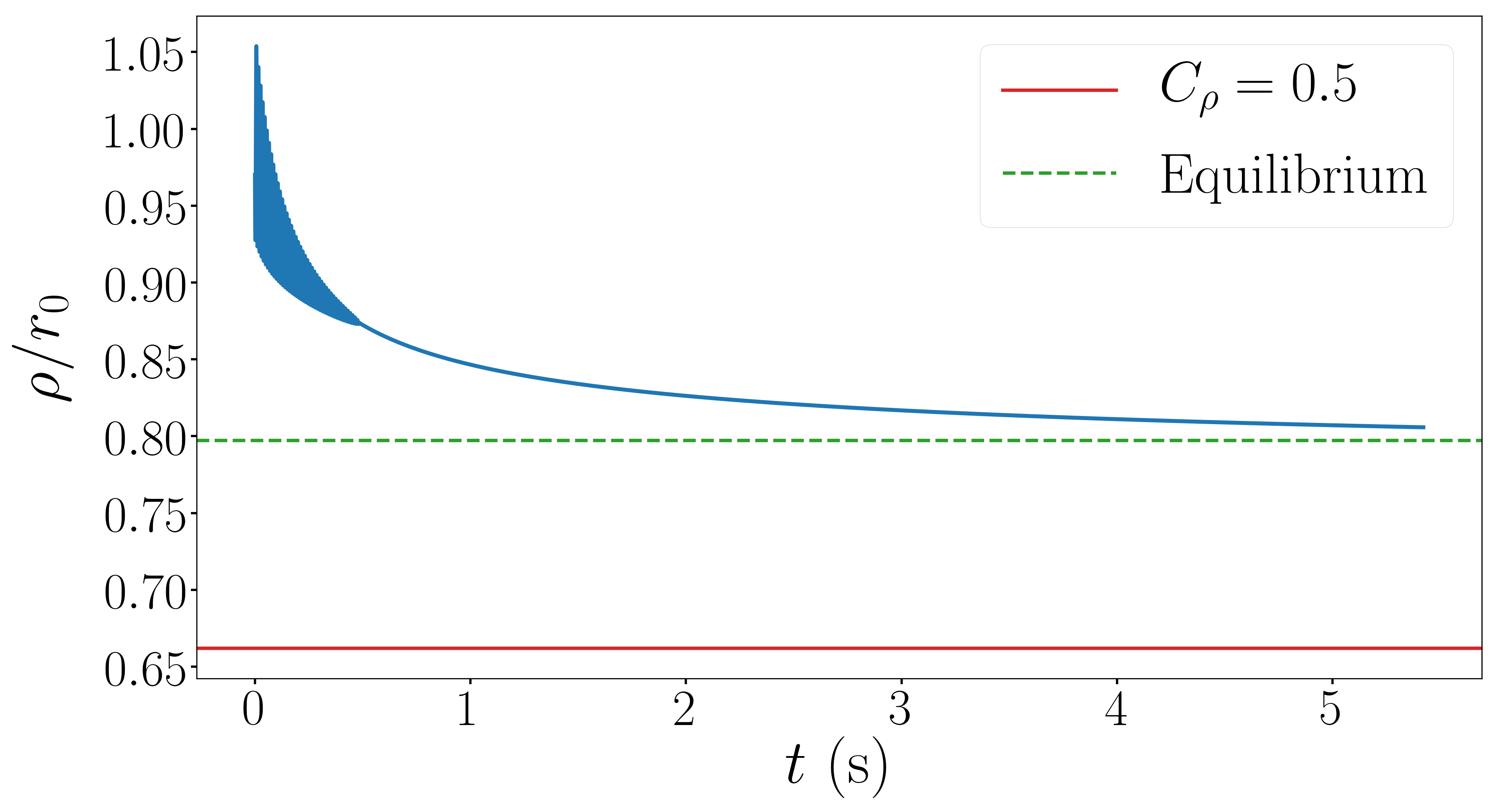}
    }
    \hspace{2em}
\subfigure[The orbital angular velocity first oscillates around a decreasing mean value and once the orbit ``circularizes'', $\omega$ decreases while tending to 0 (not shown in the figure).]{
    \centering \includegraphics[scale=0.085]{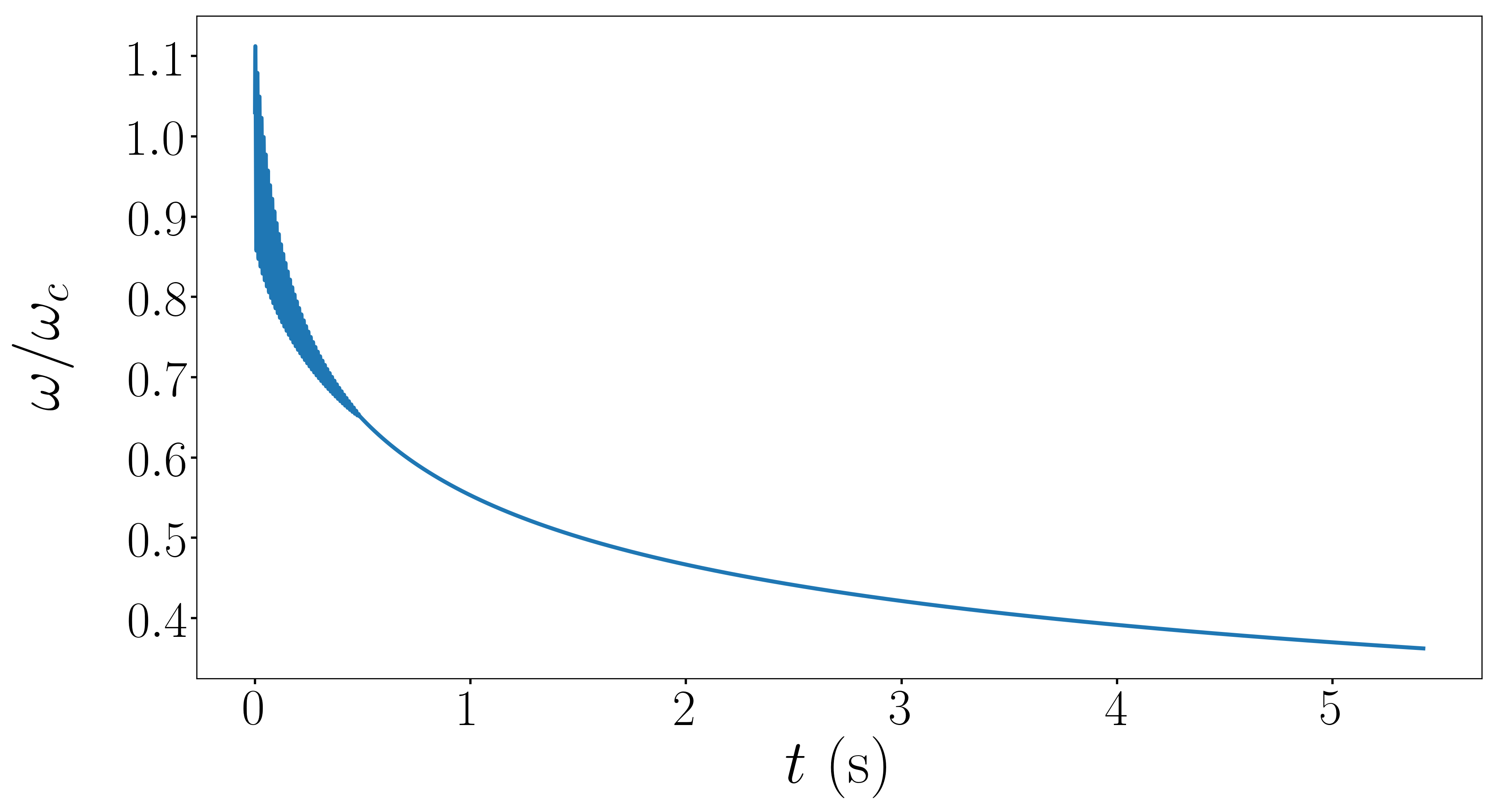}
   }
   \caption{Evolution of the distance of the cores to the center of mass (half the distance between the cores) and the orbital angular velocity for an \emph{RBH} (upper panel) and an \emph{RS} (lower panel) system.}\label{dynamics}   
 \end{figure*}

Fig.~\ref{dynamics} shows the post-contact evolution of an \emph{RBH} (upper panel) and an \emph{RS} (lower panel) system with $m_0=30 \ {M_{\odot}}$ and $C_0=0.17$. In the left panel we display the distance of the cores to the center of mass, which is half the separation between the cores, and in the right panel, the orbital angular velocity. Red solid lines indicate the threshold value of $\rho$ below which a BH is formed, i.e. such that $C_{\rho}$, defined in Eq.~\eqref{comp}, is 0.5. For the \emph{RBH} system, $\rho$ oscillates around a decreasing mean value and, eventually, reaches the threshold value, leading to the formation of a BH. $\omega$ oscillates around a stable mean value and increases right before the formation of the BH. This is related to the increase in the GW frequency before the ringdown stage in Fig.~\ref{wvf_td_bh}. For the system we show, the final BH has total mass $M_f=58 \ M_{\odot}$ and dimensionless spin $a_f=0.39$. 
In the case of the \emph{RS} system, $\rho$ and $\omega$ initially oscillate around a decreasing mean value. The orbit then ``circularizes'', from which point both decrease, while the system tends to an equilibrium configuration. The equilibrium value of $\rho$, indicated by the green dashed line, is higher than the threshold value for collapse. Therefore, a stable ECO is formed. The equilbrium value of $\omega$ is 0, this cannot be seen from Fig.~\ref{dynamics} because we do not display the long-term evolution for sake of clarity. The evolution of $\rho$ for \emph{NRS} systems is similar to the one of the \emph{RS} system.

\FloatBarrier

\bibliographystyle{unsrt}
\bibliography{bhm.bib}

\end{document}